\documentclass[a4paper,12pt]{article}

\pdfoutput=1

\usepackage[T1]{fontenc}
\usepackage[utf8]{inputenc}
\usepackage[english]{babel}
\usepackage{amsmath}
\usepackage{amsthm}
\usepackage{amssymb}
\usepackage{mathrsfs}
\usepackage{bm}
\usepackage{braket}
\usepackage{bbm}
\usepackage{tocloft}
\usepackage{mathtools}
\usepackage{ifthen}
\usepackage{slashed}
\usepackage{graphicx}
\usepackage{epstopdf}       
\usepackage{xspace}
\usepackage{microtype}
\usepackage{booktabs}
\usepackage{array, tabularx}
\usepackage{chngpage}
\usepackage[english]{varioref}
\usepackage[nottoc]{tocbibind}
\usepackage[numbers,sort&compress]{natbib}
\usepackage{url}
\usepackage{multicol, multirow}
\usepackage{mparhack}
\usepackage{emptypage}
\usepackage{setspace}
\usepackage{dsfont}
\usepackage{rotating}
\usepackage{enumitem}
\usepackage{titlesec}
\usepackage{comment}
\usepackage{tikz,tikz-cd}
\usepackage{xcolor}

\usetikzlibrary{cd,shadings}

\numberwithin{equation}{section}  

\usepackage[left=2.5cm, right=2.5cm, top=2.5cm, bottom=2.5cm]{geometry}
\usepackage[colorlinks=true
,urlcolor=blue
,anchorcolor=blue
,citecolor=blue
,filecolor=blue
,linkcolor=blue
,menucolor=blue
,linktocpage=true
]{hyperref}

\newlength{\bibitemsep}
\newlength{\bibparskip}
\let\oldthebibliography\thebibliography
\renewcommand\thebibliography[1]{%
  \oldthebibliography{#1}%
  \setlength{\parskip}{\bibitemsep}%
  \setlength{\itemsep}{\bibparskip}%
}

\renewcommand{\tilde}{\widetilde}

\renewcommand{\Re}{\operatorname{Re}}
\renewcommand{\Im}{\operatorname{Im}}
\DeclareMathOperator{\tr}{tr}

\DeclareMathOperator{\Ric}{Ric}

\DeclareMathOperator{\sgn}{sgn}
\DeclareMathAlphabet{\mathbfsf}{OT1}{cmss}{bx}{n}
\newcommand{\hooklongrightarrow}{\lhook\joinrel\longrightarrow}

\newcommand{\Z}{\mathbb{Z}}
\newcommand{\C}{\mathbb{C}}
\newcommand{\R}{\mathbb{R}}

\newcommand{\mb}[1]{\mathbf{#1}}
\newcommand{\mc}[1]{\mathcal{#1}}

\newcommand{\mf}[1]{\mathfrak{#1}}

\newcommand{\cA}{\mathcal{A}}

\newcommand{\cF}{\mathcal{F}}

\newcommand{\cL}{\mathcal{L}}
\newcommand{\cM}{\mathcal M}
\newcommand{\cN}{\mathcal{N}}
\newcommand{\cO}{\mathcal{O}}

\newcommand{\beq}{\begin{equation}}
\newcommand{\eeq}{\end{equation}}

\DeclareMathOperator{\Vol}{\mathrm{Vol}}

\newcommand{\hook}{\mathbin{\rule[.2ex]{.4em}{.03em}\rule[.2ex]{.03em}{.9ex}}}

\newcommand{\ii}{{\rm i}}
\newcommand{\e}{{\rm e}}
\newcommand{\E}{{\rm E}}

\newcommand{\g}{\mf{g}}
\newcommand{\rd}{{\rm d}}

\newcommand{\abs}[1]{\left| #1 \right|}
\newcommand{\zbar}{\overline{z}}
\newcommand{\vol}{{\rm vol}}
\newcommand{\ph}[1]{\phantom{#1}}
\newcommand{\identity}{\mathbbm{1}}
\renewcommand{\j}{\varphi}

\newcommand{\internal}{\Sigma_3}
\newcommand{\bulk}{Y_4}
\newcommand{\bdry}{M_3}

\titleformat{\section}{\bfseries}{\thesection.}{4pt}{}
\titlespacing{\section}{0pt}{20pt}{6pt}

\titleformat{\subsection}{\normalfont\itshape}{\thesubsection.}{4pt}{}
\titlespacing{\subsection}{0pt}{15pt}{6pt}

\titleformat{\subsubsection}{\normalfont\itshape}{\thesubsubsection.}{4pt}{}
\titlespacing{\subsubsection}{0pt}{15pt}{6pt}

\titleformat{\paragraph}{\normalfont\itshape}{\theparagraph.}{4pt}{}
\titlespacing{\paragraph}{0pt}{15pt}{6pt}

\allowdisplaybreaks %allows equations to be split between pages

\DeclareFontShape{OT1}{cmr}{mx}{n}%
{<->cmr10}{}
\newcommand{\mytitlefont}{\fontseries{mx}\selectfont}
\DeclareMathAlphabet{\titlemath}{OT1}{cmr}{mx}{n}

\newcommand{\balpha}{{\color{blue}\alpha}}

\begin{document} 

\begin{titlepage}
		
\begin{center}
		
~\\[2cm]
		
{\fontsize{29pt}{0pt} \mytitlefont Wrapped $M5$-branes\\[0.15cm] and complex saddle points}
		
~\\[1.25cm]

Pietro Benetti Genolini
\hskip1pt

~\\[0.5cm]

{\it Department of Applied Mathematics and Theoretical Physics, \\
University of Cambridge, Wilberforce Road, Cambridge, CB3 OWA, UK}

\vspace{0.3cm}

{\it Department of Mathematics, \\
King’s College London, Strand, WC2R 2LS, UK}

~\\[1.25cm]
			
\end{center}

\vspace{4.5cm}
			
\noindent 
We study the effects of the introduction of a $\vartheta$ term in minimal gauged supergravity in four dimensions. We show why this term is not present in supergravity duals of field theories arising on wrapped $M2$-branes, but is there in the case of $M5$-branes wrapping hyperbolic manifolds $\Sigma_3$, and compute the higher-derivative corrections. Having proved that the on-shell supergravity action of any supersymmetric solution can be expressed in terms of data from the fixed points of a Killing vector, we show that it is proportional to a complex topological invariant of $\Sigma_3$. This is consistent with the characteristics of the dual three-dimensional $\mathcal{N}=2$ SCFT predicted by the $3d$-$3d$ correspondence, and we match the large $N$ limit of its partition functions in the known cases.

\vfill 
	
\begin{flushleft}
October 2021
\end{flushleft}
	
\end{titlepage}

\setcounter{tocdepth}{3}
\renewcommand{\cfttoctitlefont}{\large\bfseries}
\renewcommand{\cftsecaftersnum}{.}
\renewcommand{\cftsubsecaftersnum}{.}
\renewcommand{\cftsubsubsecaftersnum}{.}
\renewcommand{\cftdotsep}{6}
\renewcommand\contentsname{\centerline{Contents}}

\tableofcontents

\medskip

\section{Introduction and summary}
\label{sec:Intro}

Four-dimensional Einstein--Maxwell theory with a negative cosmological constant describes the bosonic sector of minimal gauged supergravity \cite{Freedman:1976aw}. Much of the recent interest in this theory is driven by the fact that its supersymmetric asymptotically locally AdS solutions $\bulk$ describe the gravity duals to three-dimensional $\cN=2$ SCFTs on curved backgrounds $\bdry$, with $\partial\bulk \cong \bdry$. The holographically renormalized on-shell action of the supergravity solution $\bulk$ is related by the AdS/CFT correspondence to a large $N$ limit of the free energy of the dual SCFT on $\bdry$. The converse problem is more involved: given a SCFT on $\bdry$ with some geometric structure, there are many $\bulk$ filling $\bdry$ with boundary conditions matching the geometry of $\bdry$. It is expected that the gravity partition function receives contribution from all of them, and that different ones would give the dominant contribution in a saddle point approximation in different regimes of the parameters.\footnote{In fact, it is known that different methods for taking the large $N$ limit in field theory lead to matching with different supergravity solutions \cite{Toldo:2017qsh}.} However, a saddle point approximation may receive leading contributions from complex on-shell actions. This fact is well-studied in the context of AdS$_5$/CFT$_4$, and an investigation of the role of the solutions contributing to the gravity integral has been recently done in \cite{Aharony:2021zkr}. Instead, we work in four Euclidean dimensions, restricting ourselves to real metric and gauge field configurations.\footnote{This is more restrictive than the supersymmetric complex geometries that have also been studied in this context. In some cases with scalars, these have been shown to be necessary in order to match observables of four-dimensional gravity and three-dimensional field theory  (see for instance \cite{Freedman:2013oja} and \cite{Bobev:2020pjk}).} Yet, as pointed out in \cite{Choi:2020baw}, it is possible to see a sign oscillation in some field theory partition functions. To match this in supergravity, we necessarily have to introduce an imaginary $\vartheta$ term for the Abelian gauge field, which by definition does not affect the equations of motion or the supersymmetry variations. 

The holographically renormalized on-shell action of any smooth supersymmetric solution can be expressed in terms of geometric data \cite{BenettiGenolini:2019jdz}. More precisely, any supersymmetric solution admits a Killing vector $\xi$ constructed from the Killing spinor. The holographically renormalized on-shell action of a supersymmetric solution can then be expressed just in terms of contributions from the fixed point set of $\xi$, composed of isolated fixed points (nuts) and surfaces (bolts). We show that the same remains true after the introduction of the $\vartheta$ term, namely we show that
\beq
\label{eq:Iintro}
\begin{split}
I &= \sum_{\mathrm{nuts}_\mp} \pm \left( \frac{\pi}{2G_4} \pm \frac{\ii\vartheta}{2} \right) \frac{(b_1\pm b_2)^2}{4b_1b_2} \\
& \ \ \ \ + \sum_{\rm bolts \ \Sigma_\mp} \left( \frac{\pi}{2G_4} \pm \frac{\ii\vartheta}{2} \right) \int_{\Sigma_\pm} \left( \frac{1}{2}c_1(T\Sigma_\pm) \pm \frac{1}{4} c_1(N\Sigma_\pm) \right) \, .
\end{split}
\eeq
Here $\pm$ refer to the chirality of the Killing spinor at the fixed point; $b_1, b_2$ are the weights of the rotations generated by $\xi$ on the tangent space to the nut; $c_1(T\Sigma)$ and $c_1(N\Sigma)$ are the first Chern classes of the tangent and normal bundles to the bolt.\\
A crucial ingredient in the renormalization of the on-shell action in presence of the $\vartheta$ term is a finite counterterm for the $U(1)$ gauge field. This is necessary in order to have consistency with the $SL(2,\Z)$ action on $3d$ $\cN=2$ SCFTs \cite{Witten:2003ya}, and with the dependence of the partition function on the background $\bdry$ \cite{Closset:2013vra}.

One of the far-reaching properties of the formula for the on-shell action is that we don't need the analytic form of the metric, which is generally quite difficult to find, but rather only knowledge of the topology of $\bulk$ and of the circle action generated by $\xi$. This poses conceptual problems, such as suggesting the existence of an underlying fixed point theorem acting on the supergravity background, especially given that this localization persists also for corrections to the two-derivative model \cite{Bobev:2021oku, Genolini:2021urf, Hristov:2021zai}, and it also applies to complex metrics \cite{Cassani:2021dwa}. More concretely, it provides a way of computing the on-shell action of any smooth solution, assuming its existence.

It is possible to trace the origin of the $\vartheta$ term back to eleven dimensions. Four-dimensional minimal supergravity may be derived by consistently truncating eleven-dimensional supergravity on seven-dimensional internal spaces corresponding to different configurations of $M$-branes \cite{Gauntlett:2007ma, Larios:2019lxq}. Arrangements of $M2$-branes lead to $3d$ worldvolume theories that are typically Chern--Simons-matter theories. The dual eleven-dimensional supergravity solution has purely electric four-form flux and is an extension of the Freund--Rubin solution where the internal space is a seven-dimensional Sasaki--Einstein manifold. We show that a four-dimensional $\vartheta$ term cannot arise from this consistent truncation.

On the other hand, there is also an eleven-dimensional solutions with an AdS$_4$ factor and magnetic four-form flux \cite{Gauntlett:2006ux}, which is dual to $3d$ SCFTs obtained from $M5$-branes wrapping hyperbolic three-cycles $\internal$. The seven-dimensional internal space is an $S^4$ bundle over $\internal$, and there is a consistent truncation of eleven-dimensional supergravity on this space that leads to four-dimensional minimal supergravity. We prove that in this reduction the eleven-dimensional topological term reduces to a four-dimensional $\vartheta$ term for the gauge field, confirming the statements in \cite{Choi:2020baw}. Therefore, it is necessary to include a $\vartheta$ term when studying the gravity dual of the field theories obtained by wrapping $M5$-branes on hyperbolic three-manifolds, which is a case much less studied than its previously considered counterpart \cite{Gang:2014ema, Gang:2015wya, Gang:2018hjd, Gang:2019uay, Bobev:2019zmz, Benini:2019dyp, Bobev:2020zov, Choi:2020baw}.

The equations of motion for the eleven-dimensional four-form are corrected by quantum effects, and the lowest-order correction has been determined \cite{Vafa:1995fj, Duff:1995wd}. Reducing the corresponding term in the action, we are able to find the sub-leading correction to the expression for $\vartheta$:
\beq
\label{eq:thetaintro}
\vartheta = cs(\internal) \frac{2N^3-N}{3} \, ,
\eeq
where $N$ is the number of $M5$-branes, and $cs(\internal)$ is the Chern--Simons invariant of the hyperbolic $\internal$. The expression of $G_4$ in terms of the internal geometry can be determined at leading order in the usual way and it's proportional to the volume of $\internal$. More precisely, the expressions for the two four-dimensional quantities are such that the combination appearing in \eqref{eq:Iintro} takes the form
\beq
\frac{\pi}{2G_4} - \frac{\ii\vartheta}{2} \ \longrightarrow \ \frac{N^3}{3\pi} \left( \Vol(\internal) - \pi\ii \, cs(\internal) \right) \, .
\eeq
The bracketed quantity is a complex topological invariant of $\internal$ defined modulo $2\pi^2\ii$ for compact $\internal$, the \textit{complex hyperbolic volume}. This is because Mostow's rigidity theorem guarantees that the volume is a topological invariant for hyperbolic three-manifolds. 

Therefore, \eqref{eq:Iintro} tells us that the large $N$ limit of the SCFT obtained by looking at the IR limit of $N$ $M5$-branes wrapping $\internal$, denoted by $T_N[\internal]$, is proportional to a topological invariant of $\internal$. This is to be expected by effective field theory reasoning (the hyperbolic metric is unique, and the details are washed out in the IR), but moreover is perfectly consistent with the field-theoretic constructions of the $3d$-$3d$ correspondence \cite{Terashima:2011qi, Dimofte:2011ju, Cecotti:2011iy, Dimofte:2011py}. Indeed, the latter conjectures that the partition function of $T_N[\internal]$ equals that of a $SL(N,\C)$ Chern--Simons theory on $\internal$, and this relation can be used to compute the large $N$ limit of the partition function of $T_N[\internal]$.

\medskip

\noindent
\textbf{Outline}

\noindent
In Section \ref{sec:Reduction} we show how the holographically renormalized on-shell action of any smooth supersymmetric solution can be expressed in terms of data of the circle action generated by the Killing vector guaranteed by supersymmetry, even in presence of a $\vartheta$ term. This extends the results of \cite{BenettiGenolini:2019jdz} and shows the need of a finite counterterm for the gauge field. Then, in Section \ref{sec:Examples}, we consider a number of examples, both with an analytic expression for the fields and without it, including the example considered in \cite{Choi:2020baw}.\\
In Section \ref{sec:OriginTheta} we consider two consistent truncations of eleven-dimensional supergravity. We briefly review the case of $\bulk\times SE_7$, showing how there cannot be a $\vartheta$ term. We then consider in detail an internal space locally $\internal\times S^4$, studying the reduction of seven-dimensional maximal gauged supergravity on $\internal$ and finding the presence of a $\vartheta$ term. We also include some topological considerations on the global form of the eleven-dimensional solution, and the computation of the subleading term in \eqref{eq:thetaintro}.\\
Finally, in Section \ref{sec:FieldTheory} we review the large $N$ limit of the partition function of $T_N[\internal]$ on the backgrounds $\bdry$ for which it is known, showing that they match the dual gravity computation.\\
In the appendices we review some conventions for the Chern--Simons terms and the construction of the complex hyperbolic volume in terms of integral of an $SL(2,\C)$ connection, and the Bott--Cattaneo formula needed for the supergravity reductions.

\section{Reduction of the supergravity action}
\label{sec:Reduction}

\subsection{Action}
\label{subsec:Action}

We consider four-dimensional Einstein--Maxwell theory with a cosmological constant and a $\vartheta$ term. The action describes the metric $g$ and the gauge field $A$ with curvature $F=\rd A$ with action
\beq
\label{eq:4dSUGRAAction}
S = - \frac{1}{16\pi G_4}\int \left( R_g + 6 - F^2 \right) \vol_g + \frac{\ii\vartheta}{8\pi^2}\int F \wedge F \, .
\eeq
The equations of motion coming from this action are
\beq
\label{eq:4dEOMs}
\begin{split}
0 &= (\Ric_g)_{\mu\nu} + 3 g_{\mu\nu} - 2\left( F_{\mu\rho}F_\nu^{\ph{\nu}\rho} - \tfrac{1}{4} F^2 g_{\mu\nu} \right) \, , \\
0 &= \rd *_g F \, .
\end{split}
\eeq
With these conventions, the vacuum solution is given by AdS$_4$ with unit radius, that is, the metric is normalized to have constant sectional curvature equal to $-1$.

\medskip

The action \eqref{eq:4dSUGRAAction} describes the bosonic part of the minimal gauged supergravity in four dimensions. A classical solution is supersymmetric if there is a non-vanishing Dirac spinor $\epsilon$ satisfying the Killing spinor equation
\beq
\label{eq:KSE4d}
0 = \left[ \nabla - \ii A_\mu + \frac{1}{2} \Gamma_\mu + \frac{\ii}{4}F_{\nu\rho}\Gamma^{\nu\rho}\Gamma_\mu \right] \epsilon \, ,
\eeq
where $\Gamma_\mu$ generate Cliff$(4,0)$. As shown in \cite{Dunajski:2010uv}, a spinor $\epsilon$ satisfying the Killing spinor equation cannot be everywhere chiral, so generically we have a non-zero Dirac spinor, which in four dimensions defines an identity structure. Concretely, in addition to a local orthonormal frame $\{ \E^1, \E^2, \E^3, \E^4 \}$, we have two functions $S$ and $\theta$
\beq
S \equiv \epsilon^\dagger \epsilon \, , \qquad \cos^2 \frac{\theta}{2} \equiv \frac{\epsilon^\dagger_+ \epsilon_+}{S} \, , \qquad \sin^2 \frac{\theta}{2} \equiv \frac{\epsilon^\dagger_- \epsilon_-}{S}
\eeq
where $\epsilon_\pm \equiv \frac{1}{2}\left( \identity \pm \Gamma_* \right) \epsilon$ are the chiral projections of the spinor. The structure degenerates at the points where the spinor vanishes and where it becomes chiral.\footnote{\label{footnote:spinc}Since $\epsilon$ has unit charge under the $U(1)$ gauge field, it is generically a section of a spin$^c$ bundle over the manifold $M$, and thus the local frame is twisted. For more details, we refer to \cite{Martelli:2012sz, BenettiGenolini:2019jdz}.}

Using the bilinears constructed from the spinor and making up the identity structure, the Killing spinor equation \eqref{eq:KSE4d} can be shown to imply a number of differential equations equivalent to the equations of motion \eqref{eq:4dEOMs}. In particular, supersymmetric solutions are naturally provided with a $U(1)$ isometry generated by a bilinear vector field 
\beq
\xi^\flat \equiv - \ii \epsilon^\dagger \Gamma_{(1)}\Gamma_* \epsilon
\eeq 
that is also a symmetry of all the bosonic fields:
\beq
\cL_\xi S = \cL_\xi \theta = 0 \, , \qquad \cL_\xi F = 0 \, , \qquad \cL_\xi g = 0 \, .
\eeq
In fact, we have
\beq
\label{eq:dxib}
\rd\xi^\flat = -2 S \left( \E^{34} + \cos\theta \, \E^{12} + *_g F + \cos\theta \, F \right)
\eeq

\medskip

Because of the results of the supersymmetry analysis, we focus on four-dimensional space-times solving the equations \eqref{eq:4dEOMs} that admit a $U(1)$ symmetry generated by the vector field $\xi$ preserving both metric and gauge field.\footnote{\label{footnote:Orbits}More precisely, the action is $U(1)$ only if all orbits close, which we assume. If there are non-closed orbits, we would still have a torus action, and then we could approximate $\xi$ by a sequence of Killing vectors of the $U(1)^2$. For more details, we refer to \cite{BenettiGenolini:2019jdz}.} This assumption means that we consider manifolds $\bulk$ that have the structure of a circle fibration over an orbifold base $B$
\beq
\label{eq:S1Fibration}
S^1 \hooklongrightarrow \bulk \setminus (\bulk)_0 \longrightarrow B \, ,
\eeq
where $(\bulk)_0$ is the subset of fixed points for the $U(1)$ symmetry generated by $\xi$. Upon introducing an angle coordinate $\psi \sim \psi + \beta$ adapted to $\xi$, we may write the line element for $g$ on $\bulk\setminus (\bulk)_0$ as
\beq
\rd s^2 = V ( \rd \psi + \phi)^2 + V^{-1} \gamma_{ij} \rd x^i \rd x^j \, .
\eeq
Here we have also introduced a local one-form $\phi$ with $\cL_\xi \phi = 0$ and $\xi \hook \phi = 0$, a metric $\gamma$ on $B$, and the square norm of the Killing vector $V = \langle \xi , \xi \rangle_g$. It is a strictly positive global function on $B$, so $V^{-1}$ is well-defined, and we use it to rescale
\beq
\eta \equiv V^{-1}\xi^\flat \, .
\eeq
This is a global one-form on $B$ whose local expression is $\rd\psi + \phi$. Because of our assumptions on the symmetry of the spacetime, locally we can use gauge redundancy to write the following expression for the gauge field
\beq
A = \j \, \eta + a \, , \qquad F = \rd\j \wedge \eta + \j \, \rd\eta +f \, ,
\eeq
where $a$ is a transverse one form, $\xi \hook a = 0$, $f=\rd a$, and $\j$ and $a$ are both invariant under the vector field $\xi$. A constant gauge transformation along the $\rd\psi$ direction results in 
\beq
\label{eq:GaugeTransformation}
\j \mapsto \j + c \, , \qquad f \to f - c \, \rd\eta \, .
\eeq

Our purpose is to find an expression for the on-shell action that is exact on $B$. To do so, we reduce the action \eqref{eq:4dSUGRAAction} on the circle fiber, and look at the resulting equations of motion, which correspond to the reduction of \eqref{eq:4dEOMs}. Since the $\vartheta$ term does not modify the equations of motion, we may borrow the discussion from \cite{BenettiGenolini:2019jdz}. The result is that the bulk on-shell action may be written as an integral over the base $B$ as
\beq
\label{eq:Ibulk}
I_{\rm bulk} = \int_B \rd \left[ \frac{\beta}{16\pi G_4} *_\gamma \left( \rd \log V - \sigma H + 4V^{-1}\j \, \rd\j \right) + \frac{\ii\beta\vartheta}{8\pi^2} \left( - \j^2 \rd\eta - 2 \j f \right) \right] \, .
\eeq
Here we have introduced the \textit{twist} one-form $H \equiv *_\gamma \rd\eta$ and the \textit{nut potential} $\sigma$ defined by
\beq
\rd\sigma = (V^2 + 4\j^2 V) H +4\j V *_\gamma f \, .
\eeq
The crucial step in the discussion is now the fact that the on-shell action is exact on $B$ and, provided the bracketed integrand is a global two-form on $B$, we can apply Stokes' theorem obtaining an integral on the boundary of $B$. For asymptotically locally Euclidean AdS spacetimes, this is the union of the conformal boundary and the tubular neighbourhoods surrounding the fixed points of $\xi$, where the fibration \eqref{eq:S1Fibration} degenerates. As we shall see, for supersymmetric solutions the regularity assumptions hold.

It is easy to make sure that the expression for the bulk on-shell action is invariant under the gauge transformation \eqref{eq:GaugeTransformation}: the real part in \eqref{eq:Ibulk} shifts by a multiple of the equation of motion for $\j$, so on-shell is gauge-invariant, and the purely imaginary part shifts by a term that vanishes using the Bianchi identity.

\subsection{Local form of supersymmetric solutions}
\label{subsec:LocalForm}

A careful analysis of the differential equations obtained for the bilinears constructed from the spinor satisfying \eqref{eq:KSE4d} provides us with the general (local) form of the supersymmetric solution to \eqref{eq:4dEOMs} \cite{Dunajski:2010uv, BenettiGenolini:2019jdz}.

We introduce on $B$ a local real coordinate $y$, a local complex coordinate $z$, and a real function $W$ so that we can write the following expressions for the metric and gauge field
\begin{align}
\label{eq:LocalSUSYMetric}
\rd s^2 &= S^2\sin^2\theta (\rd\psi + \phi)^2 + \frac{1}{y^4S^2\sin^2\theta}\left(\rd y^2 + 4\e^W \rd z \rd \zbar\right) \, , \\
\label{eq:LocalSUSYGaugeField}
A &= \left( S \cos\theta + c_{\j}\right) (\rd\psi + \phi) + \frac{\ii}{4}\left( \partial_{z}W \, \rd z - \partial_{\zbar} W\, \rd \zbar\right) \, ,
\end{align}
where $W$ is costrained by
\beq
\label{eq:FirstConstraint}
\frac{y}{4}\partial_y W = 1 - \frac{1}{yS \sin^2\theta} \, .
\eeq
These local expressions immediately give the following
\beq
\label{eq:SUSYQuantities}
V = S^2 \sin^2\theta \, , \qquad \gamma = \frac{1}{y^4}\left(\rd y^2 + 4\e^W \rd z \rd \zbar\right) \, .
\eeq
We notice that the points of $\bulk$ where $V=0$ correspond to the points where $S=0$ or $\theta = 0,\pi$, which is where the spinor vanishes or becomes chiral. Therefore, we see that $(\bulk)_0$, the fixed points of the isometry, are also the points where the supersymmetry-induced identity structure degenerates.

Most importantly, we find expressions for $\j$ and the nut potential $\sigma$
\beq
\label{eq:jandsigma}
\j = S \cos\theta + c_\j \, , \qquad \sigma = 2S\j + c_\sigma \, ,
\eeq
which guarantee that $\j$ and $\sigma$ are both global regular functions on $(\bulk)_0$, and the constants $c_\j, c_\sigma$ will be fixed later on. For completeness, we also write down the expressions for $\rd\eta$
\beq
\label{eq:SecondConstraint}
\rd\eta = 2V^{-3/2} *_{\gamma} \left[ 2\cot\theta \, \rd\left(\frac{1}{y}\right) - S\, \rd\theta\right] \, ,
\eeq
and two additional constraints on the functions coming from higher order differential forms constructed from the spinor
\begin{align}
\label{eq:ThirdConstraint}
\partial^2_{z\zbar}W &= - \e^W\left[\partial^2_{yy}W + \frac{1}{4}(\partial_yW)^2 + \frac{12\cos^2\theta}{y^4S^2\sin^4\theta} \right] \, , \\
\begin{split}
\label{eq:GenericDzbarzf}
\partial^2_{z\zbar}f &+ \frac{\e^W}{y^2}\bigg[f\left(f^2+2\right) - y\left(2\partial_yf+\frac{3}{2}f\partial_y W\right)\ + \\
& + y^2\left(\partial_{yy}^2f + \frac{3}{2}\partial_yW\partial_yf + \frac{3}{2}f\partial_{yy}^2W + \frac{3}{4}f(\partial_yW)^2\right) \bigg]\, = \, 0 \ .
\end{split} 
\end{align}
In the latter equation, the function $f$ is defined by
\beq
f \equiv -\frac{2\cos\theta}{yS \sin^2\theta} \, .
\eeq

\subsection{On-shell action}
\label{subsec:OSAction}

Thanks to the analysis of the supersymmetry equations, we have showed that the bulk on-shell action \eqref{eq:Ibulk} is globally exact on $B$ once we remove the points where the fibration \eqref{eq:S1Fibration} degenerates. More formally, we remove a tubular neighbourhood of radius $\varepsilon$ around each connected component of $(\bulk)_0$ to construct $(\bulk)_\varepsilon \subset \bulk$ and project to $B_\varepsilon \equiv (\bulk\setminus (\bulk)_\varepsilon)/S^1 \subset B$, which has boundaries surrounding the orbifold points (in addition to the conformal boundary). The two-form in \eqref{eq:Ibulk} is globally defined on $B_\varepsilon$, so we may use Stokes' theorem and rewrite it as a sum of contributions from the UV conformal boundary and from the boundaries surrounding the fixed points of the isometry.

\medskip

Starting from the former, for asymptotically locally Euclidean AdS solutions we interpret $1/y$ as a radial coordinate near the conformal boundary, which we set at $\{y=0\}=\partial \bulk \cong \bdry$, and we assume that the various functions appearing the metric and gauge field have an analytic expansion in $y$ near the boundary. We then cut-off the spacetime to $Y_\delta$ where $y$ decreases up to $y=\delta>0$. The component of the boundary $\partial Y_\delta \equiv \{ y=\delta \} \cap \bulk \cong \bdry$ is still a fibered space with base $\partial B_\delta \equiv \partial Y_\delta /S^1$. The bulk on-shell action \eqref{eq:Ibulk} gives a contribution from the UV boundary $\partial B_\delta$ that can be lifted to an integral over $\partial Y_\delta$ with the form
\beq
I^{\rm UV}_{\rm bulk} = - \int_{\partial Y_\delta} \eta \wedge \left[ \frac{\beta}{16\pi G_4} *_\gamma \left( \rd \log V - \sigma H + 4V^{-1}\j \, \rd\j \right) - \frac{\ii\beta\vartheta}{8\pi^2} \left( \j^2 \rd\eta + 2 \j f \right) \right]
\eeq
where the sign is due to the relation between the orientation imposed by supersymmetry and that required to apply Stokes' theorem. We should also include the Gibbons--Hawking--York term and the standard counterterms implementing the holographic renormalization \cite{Emparan:1999pm, Skenderis:2002wp}, which are constructed using the metric $h$ induced on the hypersurface $\partial Y_\delta$
\begin{align}
I_{\rm GHY} &= - \frac{1}{8\pi G_4}\int_{\partial Y_\delta} K \, \vol_h \, , \\
I_{\rm ct} &= \frac{1}{8\pi G_4} \int_{\partial Y_\delta} \left( 2 + \frac{1}{2} R_h \right) \vol_h \, .
\end{align}

\medskip

In order to evaluate them, we impose the constraints \eqref{eq:FirstConstraint},  \eqref{eq:SecondConstraint}, \eqref{eq:ThirdConstraint} and \eqref{eq:GenericDzbarzf} in the expansions of $\theta$, $W$, $S$ and $\phi$, obtaining
\begin{align}
\theta &= \frac{\pi}{2} + y\, \theta_{(1)} + \frac{y^2}{2}\theta_{(2)} + \mc{O}(y^3) \, , \\
W &= W_{(0)} + y\, W_{(1)} + \frac{y^2}{2}\left( -\e^{-W_{(0)}}\partial_{z,\zbar}^2W_{(0)} - 12\theta_{(1)}^2 - \frac{1}{4}W_{(1)}^2 \right) + \mc{O}(y^3) \, , \\
\begin{split}
S &= \frac{1}{y} + \frac{1}{4}W_{(1)} + y \, \left( - \frac{1}{4}\e^{-W_{(0)}}\partial_{z,\zbar}W_{(0)} - 2\theta_{(1)}^2 \right) + \frac{y^2}{2}\bigg[ \frac{1}{8} e^{-W_{(0)}} \Big(\partial^2_{z,\zbar}W_{(0)}W_{(1)}  \\
& \ \ \ \ - 2 \partial^2_{z,\zbar}W_{(1)}\Big)  + \frac{1}{2} W_{(1)}\theta_{(1)}^2 - \theta_{(1)}\theta_{(2)} \bigg] + \mc{O}(y^3) \, ,
\end{split} \\
\phi &= \phi_{(0)} + y^2\, \ii ( \partial_{\zbar}\theta_{(1)}\, \rd \zbar - \partial_{z}\theta_{(1)}\, \rd z ) + \mc{O} ( y^3) \, ,
\end{align}
where $\phi_{(0)}$ is constrained to satisfy
\beq
\label{eq:dphi0}
\rd\phi_{(0)} = 4\ii \e^{W_{(0)}}\theta_{(1)}\, \rd z \wedge \rd \zbar \, .
\eeq
The conclusion, as detailed in \cite{Genolini:2016ecx} is that the boundary data $W_{(0)}, \theta_{(1)}$, together with the free bulk functions $W_{(1)}, \theta_{(2)}$, determine the terms at the higher orders of the expansion.

\medskip

Expanding \eqref{eq:LocalSUSYMetric} and \eqref{eq:LocalSUSYGaugeField} near $\partial \bulk$, we confirm that we have an asymptotically locally AdS space and on the conformal boundary $\bdry$ we may pick the following representatives for the metric and gauge field
\begin{align}
\label{eq:BoundaryMetric}
\rd s^2_{(0)} &= \eta_{(0)}^2 + 4 \e^{W_{(0)}} \rd z \rd \zbar \, ,  \\
\label{eq:BoundaryGaugeField}
A_{(0)} &= - \theta_{(1)} \, \eta_{(0)} + \frac{\ii}{4} \left( \partial_z W_{(0)} \, \rd z - \partial_{\zbar}W_{(0)} \, \rd \zbar \right) \, , 
\end{align}
where $\eta_{(0)} = \rd\psi + \phi_{(0)}$, $\vol_2 = 2\ii \e^{W_{(0)}} \rd z \wedge \rd \zbar$ and $\square \equiv \e^{-W_{(0)}}\partial^2_{z,\zbar}$. This is the same geometry imposed by the presence on a three-dimensional space of two supercharges with opposite $R$-charge \cite{Closset:2012ru}. 

Indeed, this can be seen explicitly by analysing the Killing spinor equation \eqref{eq:KSE4d} near the boundary, where it reduces to a conformal Killing spinor equation for a two-component spinor $\zeta$, the Killing spinor equation for three-dimensional conformal supergravity \cite{Klare:2012gn}. If $\zeta$ is nowhere vanishing, this is in turn equivalent to a solution of the Killing spinor equation of three-dimensional new minimal supergravity. The second ``supercharge'' on the three-manifold is the charge conjugate spinor $\zeta^c \equiv \mc{C}^{-1}\zeta^*$, which satisfies the Killing spinor equation with opposite $R$-charge. Without going into a detailed analysis of the possible backgrounds (see \cite{Closset:2019hyt} for a review), in our case from the two supercharges we can construct a real Killing vector that generates a transversely holomorphic foliation. This is precisely the restriction to $\bdry$ of $\partial_\psi$, and the almost contact structure one-form is $\eta_{(0)}$. By assumption, all the orbits of $\xi$ close in the bulk (see footnote \ref{footnote:Orbits}), and so do the orbits of $\xi|_{\bdry}$, so the geometry of $\bdry$ is that of a Seifert manifold, that is, a circle bundle over a two-dimensional orbifold.

\medskip

The non-divergent UV contribution to the on-shell action is given by
\beq
\begin{split}
I^{\rm UV} &= \lim_{\delta \to 0} \left[ I^{\rm UV}_{\rm bulk} + I_{\rm GHY} + I_{\rm ct} \right] \\
&= \frac{1}{16\pi^2 G_4} \int_{\bdry}\eta_{(0)}\wedge \vol_2  \Big[ 2\pi c_\sigma \theta_{(1)} \\
& \qquad \qquad \qquad \qquad + c_\j \left( -\ii G_4 \vartheta \square W_{(0)} + 2\pi \theta_{(2)} + \theta_{(1)} \left(- 4\ii c_\j G_4\vartheta + 3\pi W_{(2)} \right) \right)\\
& \qquad \qquad \qquad \qquad + \ii G_4 \vartheta \, \left( \theta_{(1)} \, \square W_{(0)} + 4 \theta^3_{(1)} \right) \Big] \, .
\end{split}
\eeq
There is a natural choice that simplifies the UV contribution to the on-shell action, namely $c_\sigma = c_\j = 0$, which was called \textit{supersymmetric gauge choice} in \cite{BenettiGenolini:2019jdz}
\beq
\label{eq:IOS_UV}
\begin{split}
I^{\rm UV} &= \frac{\ii \vartheta }{32\pi^2} \int_{\partial M}\eta_{(0)}\wedge \rd\eta_{(0)} \left( - R_{2d} + (*_2\rd\eta_{(0)})^2 \right) \, ,
\end{split}
\eeq
where $R_{2d}\equiv - \square W_{(0)}$.
Notice that this is a well-defined integral over $\partial M$, and that the only UV contribution to the on-shell action comes from the $\vartheta$ term, whereas the ``standard'' gravitational part is determined only be the fixed points of the isometry.

\medskip

We could also consider the inclusion of additional finite counterterms constructed out of the gauge fields. For standard Maxwell theory, no such counterterm is required if one fixes at the gauge field at the boundary, imposing that the dual operator in the CFT is a $U(1)$ current, as we do \cite{Marolf:2006nd}. However, in this case we also have a non-zero $\vartheta$ term, which could make the issue a bit subtler. We suggest the necessity of the following counterterm
\beq
\label{eq:FiniteCounterterm}
I_{{\rm ct}, A} = - \frac{\ii\vartheta}{2\pi}I_{CS}[A_{(0)}] \, ,
\eeq
where $I_{CS}[A]$ is the Chern--Simons action at level one, which for a trivial $U(1)$ bundle can be written as
\beq
\label{eq:ChernSimonsLevelOne}
I_{CS}[A] = \frac{1}{4\pi}\int_{M_3} A \wedge \rd A \, .
\eeq
The concrete effect of this counterterm is cancelling $I^{\rm UV}$. We shall discuss this in more detail in Section \ref{subsec:BdryTheta}.

\medskip

We now move to the evaluation of the contributions to the on-shell action from the other components of the boundary of $B_\varepsilon$, namely the tubular neighbourhoods surrounding the fixed points of the isometry acting on $\bulk$. These must have even codimension, so in four dimensions we may split them into zero-dimensional nuts and two-dimensional bolts, following the terminology of \cite{Gibbons:1979xm}. As we noticed around \eqref{eq:SUSYQuantities}, the fixed points of the isometry are also the loci where the identity structure degenerates. However, a more careful analysis of the bilinear equation \eqref{eq:dxib} shows that $S$ may never vanish, unless we are in the trivial case where the isometry acts trivially everywhere. Therefore, the only possibility for the fixed points is that the spinor becomes chiral there. This allows us to further split the classification of the connected components of the fixed point sets according to the chirality of the supersymmetry spinor: we have nuts$_\pm$ and bolts $\Sigma_\pm$ depending on whether $\epsilon$ has positive/negative chirality there. Furthermore, \eqref{eq:dxib} also gives us that at a fixed point with chirality $\pm$
\beq
\label{eq:S2pm}
S^2\rvert_\pm = \frac{1}{16} \langle (\rd \xi^\flat\rvert_\pm )^\mp , (\rd \xi^\flat\rvert_\pm )^\mp \rangle_g \, ,
\eeq
where the superscript $\pm$ indicates the (anti-)self-dual part of the two-form.\\
We split the contributions of the bulk into the ``standard'' part, studied in \cite{BenettiGenolini:2019jdz}, and the $\vartheta$ term: from \eqref{eq:Ibulk} we write
\beq
\label{eq:IbulkIR}
\begin{split}
I_{\rm bulk}^{\rm IR} &= - \int_{\partial B_\varepsilon} \left[ \frac{\beta}{16\pi G_4} *_\gamma \left( \rd \log V - \sigma H + 4V^{-1}\j \, \rd\j \right) + \frac{\ii\beta\vartheta}{8\pi^2} \left( - \j^2 \rd\eta - 2 \j f \right) \right] \\
&\equiv I^{\rm IR}_0 + I^{\rm IR}_{\vartheta} \, ,
\end{split}
\eeq
where
\beq
\label{eq:IthetaIR}
I^{\rm IR}_{\vartheta} = \frac{\ii\vartheta}{8\pi^2}\int_{\partial M_\varepsilon} \eta \wedge \j \left( - \j \, \rd\eta + 2F \right) \, .
\eeq

\medskip

In a neighbourhood of a nut, we can write the Killing vector as the generator of the rotations on the two two-planes of $\R^4 \cong \R^2 \oplus \R^2$ with weights $b_1, b_2$, that is
\beq
\xi = b_1 \partial_{\j_1} + b_2 \partial_{\j_2} \, ,
\eeq
where $\j_1, \j_2$ are the polar coordinates on the two copies of $\R^2$ normalised to have $2\pi$ periodicity. Since we are assuming that the orbits of $\xi$ all close, the ratio between the two weights is rational, so we write $b_1/b_2 = p/q$ with $p,q$ coprime and the period of the generic orbit is $\beta = 2\pi p/b_1 = 2\pi q/b_2$. The connected component of $\partial B_\varepsilon$ is thus the weighted complex projective space $\mathbb{WCP}^1_{[p,|q|]}$ and so the orbifold line bundle $\cL$ associated to $S^1 \hookrightarrow \partial (\bulk)_\varepsilon \rightarrow \partial B_\varepsilon$ has first Chern class
\beq
\label{eq:detanuts}
\int_{\partial B_\epsilon}\rd\eta = \beta \int_{\mathbb{WCP}^1_{[p,|q|]}} c_1(\cL) = - \frac{\beta}{pq} \, .
\eeq
In the supersymmetric gauge for $\sigma$ and $\j$, near a nut of type $\pm$ we have $\j \rvert_\pm = \pm S \rvert_\pm$ (as can be seen from \eqref{eq:jandsigma}) and \eqref{eq:S2pm} gives
\beq
\label{eq:j2nuts}
\j^2 \rvert_{\rm nut_\pm} = \frac{1}{4} (b_1 \mp b_2)^2 \, .
\eeq
Considering \eqref{eq:IbulkIR} term by term, we can show that
\beq
I^{\rm IR}_0 \rvert_{\rm nuts} =  \frac{\pi}{2G_4}\sum_{\mathrm{nuts}_\mp} \pm \frac{(b_1\pm b_2)^2}{4b_1b_2} \, .
\eeq
As for $I^{\rm IR}_\vartheta$ in \eqref{eq:IthetaIR}, the second term in the sum is the integral over the $\partial (\bulk)_\varepsilon \cong S^3_\varepsilon$ surrounding the nut of a smooth form, so this term gives a vanishing contribution. On the other hand, the first term does contribute, and combining \eqref{eq:detanuts} and \eqref{eq:j2nuts} we have
\beq
I^{\rm IR}_{\vartheta} \rvert_{\rm nuts} = \frac{\ii\vartheta}{2} \sum_{\mathrm{nuts}_\mp} \frac{(b_1 \pm b_2)^2}{4b_1b_2} \, .
\eeq
Thus
\beq
I_{\rm nuts} = \sum_{\mathrm{nuts}_\mp} \pm \left( \frac{\pi}{2G_4} \pm \frac{\ii\vartheta}{2} \right) \frac{(b_1\pm b_2)^2}{4b_1b_2} \, .
\eeq

\medskip

Given a bolt $\Sigma_\pm$, we denote by $T_\varepsilon \cong \Sigma_\pm$ the connected component of $\partial B_\epsilon$ surrounding it. We notice that $\rd\j = - \xi \hook F$, which by regularity of $F$ implies that $\j$ is constant on the two-dimensional bolt. Therefore, we can write
\beq
\lim_{\varepsilon\to 0} \int_{T_\varepsilon} \j \, F = 2\pi \, \j \rvert_{\Sigma_\pm} \int_{\Sigma_\pm}c_1(F) \, .
\eeq
Analogously, we have
\beq
\lim_{\varepsilon\to 0} \int_{T_\varepsilon} \j^2 \rd\eta = \beta \, \j^2 \rvert_{\Sigma_\pm} \int_{\Sigma_\pm}c_1(N\Sigma_\pm) \, .
\eeq
To find the value of $\j$ at the bolt, we resort again to \eqref{eq:jandsigma} and \eqref{eq:S2pm}:
\beq
\j \rvert_{\Sigma_\pm} = \pm \frac{\pi}{\beta} \, .
\eeq
Overall, \eqref{eq:IthetaIR} gives
\beq
\label{eq:Ithetabolt_1}
I^{\rm IR}_\vartheta \rvert_{\rm bolts} = \frac{\ii\vartheta}{2} \sum_{\rm bolts \ \Sigma_{\pm}} \int_{\Sigma_\pm} \left( - \frac{1}{4} c_1(N\Sigma_\pm) \pm c_1(F) \right) \, .
\eeq
This expression can be further simplified, because the existence of supersymmetry on a curved surface imposes a sort of twist. As mentioned in footnote \ref{footnote:spinc}, the supersymmetry spinor $\epsilon$ is generically only a section of a spin$^c$ bundle that is twisted by the $U(1)$ gauge field $A$. Near a bolt, the spinor$^c$ becomes chiral but does not vanish, so there is a non-zero section of a line bundle over the bolt, meaning that said bundle must be trivial. This constraint imposes a relation between the restriction of the gauge bundle over $\Sigma_\pm$ and the tangent and normal bundle to the surface \cite{BenettiGenolini:2019jdz}
\beq
\int_{\Sigma_\pm} c_1(F) = \int_{\Sigma_\pm} \frac{\pm c_1(N\Sigma_\pm) - c_1(T\Sigma_\pm)}{2} \, .
\eeq
Inserting the latter result in the expression \eqref{eq:Ithetabolt_1} leads to the result
\beq
I^{\rm IR}_\vartheta \rvert_{\rm bolts} = \frac{\ii\vartheta}{2} \sum_{\rm bolts \ \Sigma_{\pm}} \int_{\Sigma_\pm} \left( \frac{1}{4} c_1(N\Sigma_\pm) \mp  \frac{1}{2}c_1(T\Sigma_\pm) \right) \, ,
\eeq
and a similar analysis for $I^{\rm IR}_0$ gives
\beq
I^{\rm IR}_0 \rvert_{\rm bolts} = \frac{\pi}{2G_4}\sum_{\rm bolts \ \Sigma_{\pm}} \mp \int_{\Sigma_{\pm}} \left( \frac{1}{4}c_1(N\Sigma_{\pm}) \mp \frac{1}{2}c_1(T\Sigma_{\pm}) \right) \, ,
\eeq
whence
\beq
I_{\rm bolts} = \sum_{\rm bolts \ \Sigma_\pm} \left( \frac{\pi}{2G_4} \mp \frac{\ii\vartheta}{2} \right) \int_{\Sigma_\pm} \left( \frac{1}{2}c_1(T\Sigma_\pm) \mp \frac{1}{4} c_1(N\Sigma_\pm) \right) \, .
\eeq
Summing the contributions from all the fixed points of the isometry, we find that the ``infrared'' contribution to the on-shell action with $\vartheta$ term is
\beq
\label{eq:IOS_IR}
\begin{split}
I^{\rm IR} &= \sum_{\mathrm{nuts}_\mp} \pm \left( \frac{\pi}{2G_4} \pm \frac{\ii\vartheta}{2} \right) \frac{(b_1\pm b_2)^2}{4b_1b_2} \\
& \ \ \ \ + \sum_{\rm bolts \ \Sigma_\mp} \left( \frac{\pi}{2G_4} \pm \frac{\ii\vartheta}{2} \right) \int_{\Sigma_\pm} \left( \frac{1}{2}c_1(T\Sigma_\pm) \pm \frac{1}{4} c_1(N\Sigma_\pm) \right) \, .
\end{split}
\eeq

\subsection{Dependence on the boundary and periodicity of \texorpdfstring{$\vartheta$}{theta}}
\label{subsec:BdryTheta}

Having found an expression for the on-shell action of any smooth supergravity solution, it seems redundant to investigate its dependence on boundary data. Nonetheless, it is an additional check on our expression. We compute the one-point function of the holographic stress-energy tensor and the $U(1)_R$ current: the holographic stress-energy tensor has the same form as in \cite{Genolini:2016ecx}, whereas the $U(1)_R$ current receives an additional contribution from the $\vartheta$ term
\beq
j_\vartheta = \frac{\ii\vartheta}{4\pi^2} \left[ \left( 2 \theta_{(1)}^2 + \frac{1}{4} \square W_{(0)} \right) \eta_{(0)} - *_2 \rd \theta_{(1)} \right] \, .
\eeq
Under a change in the boundary geometry and $U(1)_R$ gauge field given by $W_{(0)} \to W_{(0)} + \delta W_{(0)}, \phi_{(0)} \to \phi_{(0)} + \delta \phi_{(0)}$, where the variations are a basic function and one-form on $\partial \bulk$, respectively, we find that the variation of the holographically renormalized on-shell action vanishes \cite{Genolini:2016ecx}, apart from the $\vartheta$ term, which gives
\beq
\delta I = \frac{\ii \vartheta}{16\pi^2}\int_{\partial M}  \eta_{(0)} \wedge \delta \left( \theta_{(1)} \square W_{(0)} \, \vol_2 + 4 \theta_{(1)}^3 \, \vol_2 \right) \, .
\eeq
Looking at this, it is clear that the non-zero variation is entirely due to \eqref{eq:IOS_UV}.

There is an expectation stemming from field theory that the partition function of a three-dimensional $\cN=2$ field theory on a background preserving at least two supercharges (as the one we find at the boundary of $\bulk$) should only depend on the background geometry via the choice of transversely holomorphic foliation \cite{Closset:2013vra}. Because of the AdS/CFT correspondence, the renormalized on-shell action of the dual gravity solution should have the same dependence on the boundary data. As proved in \cite{Genolini:2016ecx}, this holds for the canonical action of four-dimensional minimal supergravity. However, the computation just presented shows that this does not hold true when we include a $\vartheta$ term, since the specific variation we considered preserves the transversely holomorphic foliation and yet $\delta I \neq 0$. This is a first piece of evidence in favour of the finite counterterm \eqref{eq:FiniteCounterterm}, which cancels $I^{\rm UV}$ thus removing $\delta I$.\footnote{The necessity of finite counterterms in order to be preserve boundary supersymmetry is well-known in cases with scalars and additional supersymmetry, e.g. \cite{Freedman:2016yue, Gauntlett:2018vhk, Bobev:2020pjk, BenettiGenolini:2020kxj}.}

\medskip

Additional evidence comes from a more careful thought on the periodicity of the $\vartheta$ angle. Abelian gauge theories in four dimensions on a spin manifold enjoy a $SL(2,\Z)$ duality group \cite{Witten:1995gf}. The $T$ generator corresponds to the shift $\vartheta \to \vartheta + 2\pi$, whereas the $S$ generator, which does not commute with $T$, is the exchange of electric and magnetic fields.

If $\bulk$ is closed and spin, then $\int_{\bulk}F\wedge F \in 8\pi^2 \Z$, so the theory for which \eqref{eq:4dSUGRAAction} is the semi-classical effective description is invariant under $\vartheta \to \vartheta + 2\pi$, since $\e^{-S}$ would be unchanged. The main focus of the paper, though, are asymptotically locally AdS spaces, more precisely cutoff at a distance $\delta$ from the conformal boundary. On a space $\bulk$ with a boundary $\bdry$, we extend the line bundle $L$ on $\bulk$ to a line bundle $L_{(0)}$ on $\bdry$. Then, $\int_{\bulk}F\wedge F$ fails to be a multiple of $8\pi^2$ by a boundary Chern--Simons term at level one \eqref{eq:ChernSimonsLevelOne}. That is, after shifting $\vartheta \to \vartheta + 2\pi$, we find
\beq
\label{eq:ShiftThetaWithoutCounterterm}
S \to S + 2\pi \ii \, \ell + \ii I_{CS}[A_{(0)}] \, , \qquad \ell \in \Z \, .
\eeq
This shift leads to an action on the partition function of the boundary theory
\beq
Z_{\rm SCFT}[A_{(0)}; \bdry] = \e^{ - S[A; \bulk]} \to Z_{\rm SCFT}[A_{(0)}; \bdry] \, \e^{ - \ii I_{CS}[A_{(0)}]} \, .
\eeq
This is expected from field theory: three-dimensional conformal field theories with a $U(1)$ global symmetry are acted upon by a $SL(2,\Z)$ group. More precisely, after coupling the $U(1)$ current to a background gauge field $A_{(0)}$, one finds that the $T$ generator of $SL(2,\Z)$ acts on the generating functional of the correlation functions of the $U(1)$ current by shifting it by a level one Chern--Simons term for the background gauge field. This is precisely our setup, since we are describing the universal sector of three-dimensional SCFTs with $\cN=2$, which always have a $U(1)_R$ global $R$-symmetry. The $SL(2,\Z)$ group acting on $3d$ CFTs becomes in the bulk the duality group of four-dimensional $U(1)$ gauge theory on AlAdS$_4$.

Notice that the boundary $SL(2,\Z)$ group is not a duality, since its action generically modifies the $3d$ CFT. Correspondingly, the $SL(2,\Z)$ action on the Einstein--Maxwell theory on anti-de Sitter space is to be interpreted as a choice of boundary conditions for the gauge field \cite{Hawking:1995ap, Witten:2003ya, Leigh:2003ez, Zucchini:2003in, Yee:2004ju}. On the other hand, the introduction of the counterterm \eqref{eq:FiniteCounterterm} fixes the choice of boundary conditions by removing the additional boundary Chern--Simons term
\beq
S + I_{{\rm ct}, A} \to S + I_{{\rm ct}, A} + 2\pi \ii \, \ell \, , \qquad \ell \in \Z \, .
\eeq
This guarantees that the bulk theory is invariant under $\vartheta \to \vartheta + 2\pi$.

\medskip

As described after \eqref{eq:BoundaryGaugeField}, the boundary superconformal field theory is formulated on $M_3$ by coupling to the multiplet of a background three-dimensional conformal supergravity, which includes $g_{(0)}$ and $A_{(0)}$. The latter is the background gauge field coupling to the $R$-symmetry current. A Chern--Simons term for $A_{(0)}$ is related by $\cN=2$ supersymmetry to the gravitational Chern--Simons term \cite{Rocek:1985bk}, so their levels are related. The fractional part of this level is a physical quantity of the SCFT: it cannot be removed by a local gauge-invariant counterterm, and it appears as a conformally-invariant contact term in the two-point function of the stress-energy tensor \cite{Closset:2012vp}. This fractional part is captured by the counterterm \eqref{eq:FiniteCounterterm}, being $\vartheta/2\pi$ mod 1, and consistently the shift $\vartheta \to \vartheta + 2\pi$ does not modify it.

\section{Examples}
\label{sec:Examples}

\subsection{\texorpdfstring{AdS$_4$}{AdS4}}
\label{subsec:AdS4}

The simplest non-trivial solution we can consider is AdS$_4$ with an instanton \cite{Martelli:2011fu, Farquet:2014kma}. The metric and gauge field are 
\beq
\begin{split}
\rd s^2 &= \frac{\rd r^2}{r^2+1} + r^2 \left( \rd\vartheta^2 + \cos^2\vartheta \, \rd\varphi_1^2 + \sin^2\vartheta \, \rd\varphi_2^2 \right) \, , \\
A &= - \frac{\left( b_1 + b_2 \sqrt{r^2+1} \right) \rd{\j_1} + \left( b_2 + b_1 \sqrt{r^2+1} \right) \rd{\j_2} }{2\sqrt{(b_2 + b_1\sqrt{r^2+1})^2\cos^2\vartheta + (b_1 + b_2 \sqrt{r^2+1})^2 \sin^2\vartheta}} \, ,
\end{split}
\eeq
where $\vartheta \in [0,\frac{\pi}{2}]$, $\j_1, \j_2 \in [0,2\pi]$, and there is a $U(1)\times U(1)$ Killing vector field constructed from the Killing spinor
\beq
\label{eq:AdSKV}
\xi = b_1 \partial_{\j_1} + b_2 \partial_{\j_2} \, .
\eeq
This solution is only regular provided $b_1/b_2 \equiv b^2 >0$ and $b_1/b_2 = -1$. In the cases $b_1/b_2 = \pm 1$ it reduces to the trivial instanton.

The norm of the Killing vector is $V= r^2 (b_1 \cos^2\phi_1 + b_2 \sin^2\phi_2 )$, which shows that at $r=0$ there is a nut. In fact, a more careful analysis leads to the conclusion that for $b_1/b_2= b^2$ we have a nut$_-$ and for $b_1/b_2=-1$ we have a nut$_+$. In the two cases applying \eqref{eq:IOS_IR} leads to
\beq
\label{eq:IAdSIR}
\frac{b_1}{b_2} = -1 \, : \quad I^{\rm IR} = \frac{\pi}{2G_4} - \frac{\ii\vartheta}{2} \, , \qquad \quad \frac{b_1}{b_2} = b^2 \, : \quad I^{\rm IR} = \frac{1}{4} \left( \frac{\pi}{2G_4} + \frac{\ii\vartheta}{2} \right) \left( b + \frac{1}{b} \right)^2 \, .
\eeq
Instead, applying \eqref{eq:IOS_UV} gives
\beq
\label{eq:IAdSUV_v1}
\begin{split}
I^{\rm UV} = - \frac{\ii\vartheta}{2}\frac{b_1^2+b_2^2}{2b_1b_2} \, ,
\end{split}
\eeq

\medskip

The smooth solution with $b_1/b_2 = b^2$ has another interpretation: AdS$_4$ (with the non-trivial instanton) can be foliated in a non-standard way to find a conformal boundary that is isometric to a squashed $S^3$ \cite{Martelli:2011fu}. More concretely, we perform the change of coordinates
\beq
\begin{split}
r^2 &= \left( \frac{q^2}{b^2} - 1 \right) \cos^2\psi + \left( q^2b^2 - 1 \right) \sin^2\psi \, , \\ 
\cos^2\vartheta &= \frac{\left( \frac{q^2}{b^2} - 1 \right)\cos^2\psi}{\left( \frac{q^2}{b^2} - 1 \right)\cos^2\psi + \left( q^2b^2 - 1 \right) \sin^2\psi} \, ,
\end{split}
\eeq
obtaining
\beq
\label{eq:NonEasyAdS4}
\begin{split}
\rd s^2 &= \frac{q^2-f(\psi)^{-2}}{(q^2-b^2)(q^2-b^{-2})} \rd q^2 + \left( q^2 f(\psi)^2 -1 \right) \rd\psi^2 \\
& \ \ \ + \left( \frac{q^2}{b^2}-1 \right) \cos^2\psi \, \rd\j_1^2 + \left( q^2 b^2 - 1 \right) \sin^2\psi \, \rd \j_2^2 \, ,
\end{split}
\eeq
with
\beq
f(\psi)^2 \equiv \frac{1}{b^{-2}\cos^2\psi + b^2 \sin^2\psi} \, .
\eeq
The metric \eqref{eq:NonEasyAdS4} is again AdS$_4$, but now the metric at the conformal boundary at $q\to \infty$ is
\beq
\label{eq:SquashedU1U1}
\rd s^2_3 = f(\psi)^2 \, \rd\psi^2 + \frac{1}{b^2} \cos^2\psi \, \rd\varphi_1^2 + b^{2} \sin^2\psi \, \rd\varphi_2^2 \, .
\eeq
This is the metric on the squashed three-sphere preserving $U(1)\times U(1)$ symmetry and two supercharges studied in \cite{Hama:2011ea}. To further confirm the smoothness of the slicings, notice that \eqref{eq:SquashedU1U1} is related to the metric on the round sphere by a Weyl transformation: upon substituting
\beq
\cos^2\vartheta = \frac{b^{-2}\cos^2\psi}{b^{-2}\cos^2\psi + b^2\sin^2\psi} \, ,
\eeq
we find that
\beq
\rd \vartheta^2 + \cos^2\vartheta \, \rd\j_1^2 + \sin^2\vartheta \, \rd\j_2^2 = f(\psi)^2 \left( f(\psi)^2 \rd\psi^2 + b^{-2}\cos^2\psi \, \rd \j_1^2 + b^2 \sin^2\psi \, \rd \j_2^2 \right) \, .
\eeq 
In fact, the form of $f(\psi)$ is not relevant for the large $N$ limit of the localization computation of the dual field theory, as showed in \cite{Alday:2013lba}. It is sufficient that it guarantees smoothness of the metric, that is, it has the behaviour
\beq
\lim_{\psi \to 0}f(\psi) = b \, , \qquad \lim_{\psi \to \frac{\pi}{2}}f(\psi) = \frac{1}{b} \, .
\eeq
Consistently, we shall see that this holds for $I^{\rm UV}$ as well. Given the metric \eqref{eq:SquashedU1U1} and the Killing vector $\xi = b \partial_{\j_1}+ b^{-1} \partial_{\j_2}$ (which is \eqref{eq:AdSKV} when $b_2 = b^{-1}$), then $\rd\eta = 2 \vol_2 /f(\psi)$, and $R_{2d} = 4(2f(\psi) +f'(\psi) \cot 2\psi )/f(\psi)^3$ so that \eqref{eq:IOS_UV} leads to
\beq
\label{eq:IAdSUV}
I^{\rm UV} = \frac{\ii\vartheta}{2} \int_0^{\pi/2} \frac{\rd \, }{\rd\psi} \left( \frac{\cos 2\psi}{2f(\psi)^2} \right) \, \rd\psi = - \frac{\ii\vartheta}{2} \frac{b^2+b^{-2}}{2} \, ,
\eeq
consistently with \eqref{eq:IAdSUV_v1} and independently on the details of $f(\psi)$.

\subsection{Supersymmetric black holes}
\label{subsec:SUSYBH}

Another simple and yet interesting solution that we consider are the supersymmetric black holes obtained by Wick-rotation of the dyonic static solutions of \cite{Romans:1991nq, Brill:1997mf, Caldarelli:1998hg}. These are related to the solutions considered in \cite{Choi:2020baw}. They preserve $1/4$ of the supersymmetry and have the topology $\R^2 \times \Sigma_g$ for $g>1$. The local form of the metric and gauge field is
\beq
\begin{split}
\rd s^2 &= V(r) \, \rd\tau^2 + \frac{\rd r^2}{V(r)} + r^2 \, \rd s^2(\Sigma_g) \, , \qquad \quad V = -1 + \frac{\frac{1}{4}-Q^2}{r^2} + r^2 \, , \\
F &= \frac{Q}{r^2}\rd\tau \wedge \rd r + \frac{1}{2} \vol(\Sigma_g)
\end{split}
\eeq
If $Q\neq 0$, the solution is regular provided $r>r_0$ with $r_0 = \sqrt{\frac{1}{2}+\abs{Q}}$ and $\tau \sim \tau + \frac{4\pi}{\abs{V'(r_0)}}$. If $Q=0$, instead, the metric develops an infinite throat near $r = 1/\sqrt{2}$.

The Killing vector $\xi$ constructed from the Killing spinor is $\xi = \partial_\tau$. It has a bolt at $r=r_0$ if $Q \neq 0$ that goes to infinite distance as $Q\to 0$. Provided $Q\neq 0$, we have a smooth metric with a bolt$_\pm$ where the chirality of the spinor at the bolt is equal to $\sgn(Q)$. Since the fibration is trivial, equation \eqref{eq:IOS_UV} vanishes and the only contribution comes from the IR \eqref{eq:IOS_IR}
\beq
\label{eq:IOSBH}
I = \left( \frac{\pi}{2G_4} - \sgn(Q) \frac{\ii\vartheta}{2} \right) (1-g) \, .
\eeq
The only dependence on $Q$ in the action is in the $\vartheta$ term, and we will obtain different results depending on whether we take $Q \to 0^+$ or $Q\to 0^-$. The same result has been obtained in \cite{Choi:2020baw} using a different regularization of the extremal metric with $Q=0$ breaking supersymmetry.

\subsection{\texorpdfstring{$\frac{1}{4}$}{1/4}-BPS solutions}
\label{subsec:14BPS}

There are also a number of solutions preserving two real supercharges but with a non-trivial fibration. Here we consider those constructed in \cite{Martelli:2012sz, Toldo:2017qsh} and further analysed in \cite{Farquet:2014kma, Genolini:2016ecx}. They all have the form
\beq
\label{eq:MetricGaugeField14BPS}
\begin{split}
\rd s^2 &= \frac{r^2-s^2}{\Omega(r)} \rd r^2 + (r^2 -s^2) (\sigma_1^2 + \sigma_2^2) + 4s^2 \frac{\Omega(r)}{r^2-s^2} \sigma_3^2 \, , \\
A &= \frac{P(r^2+s^2)- 2Qrs}{(r^2-s^2)} \sigma_3 \, 
\end{split}
\eeq
for different choices of $\Omega(r), P,Q$ and the forms $\sigma_i$.

\medskip

There is a Einstein Taub-NUT-AdS solution on $\R^4/\Z_p$ (with an instanton) for the choices
\beq
\Omega(r) = (r-s)^2 \left[ 1 + (r-s)(r+3s) \right] \, , \qquad Q = P = - \frac{4s^2-1}{2} \, ,
\eeq
and the $\sigma_i$ are the left-invariant one-forms on $S^3$
\beq
\label{eq:LeftInvariantSphere}
\sigma_1 + \ii \sigma_2 = \e^{-\ii \tau}(\rd\vartheta + \ii \sin\vartheta \, \rd \phi) \, , \qquad \sigma_3 = \rd\tau + \cos\vartheta \, \rd\phi \, .
\eeq 
This is a self-dual metric, in the sense that both the gauge field and the Weyl tensor are self-dual. The Killing vector constructed from the Killing spinor is $\xi = \partial_\tau$, so $\eta = \sigma_3$. We use the gauge freedom in \eqref{eq:GaugeTransformation} to find that the function $\j$ satisfying the supersymmetric gauge choice is
\beq
\j = - \frac{(4s^2-1)(r-s)}{2(r+s)} - \frac{1}{2} \, .
\eeq
The conformal boundary is a squashed $S^3/\Z_p$ with metric
\beq
\rd s^2_3 = \sigma_1^2 + \sigma_2^2 + 4s^2 \sigma_3^2 \, .
\eeq
This squashed sphere preserves $SU(2)\times U(1)$ and supports two real supercharges \cite{Hama:2011ea}. In order to apply \eqref{eq:IOS_UV}, we let $\eta_{(0)} = 2s \, \sigma_3$, from which $R_{2d}=2$. Thus
\beq
\label{eq:IUV14BPS}
I^{\rm UV} = \frac{\ii\vartheta}{2} \frac{16s^4 - 8 s^2}{p} \, .
\eeq
Given the form of $\Omega$ and $Q, P$, we find that $\xi$ has a nut at $r=s$, corresponding to the origin of the $\R^4$ where we should take $\tau \sim \tau + \frac{4\pi}{p}$ for regularity. In our terminology, the origin is a nut$_-$ with $b_1 = b_2$, so \eqref{eq:IOS_IR} gives
\beq
I^{\rm IR} = \frac{\pi}{2G_4} + \frac{\ii\vartheta}{2} \, .
\eeq
The overall action with $\vartheta$ term, which is consistent with \cite{Martelli:2012sz}, is the sum of the two contributions.

\medskip

In addition to the self-dual Taub-NUT-AdS solution with topology $\R^4/\Z_p$, there is also a family of solutions with topology $\mc{O}(-p) \to \Sigma_g$ for any $\Sigma_g$ with metric with constant sectional curvature $\kappa$. In this case the metric and gauge field still have the form \eqref{eq:MetricGaugeField14BPS}, but now
\beq
\begin{split}
\Omega &= (r^2-s^2)^2+ (\kappa - 4s^2)(r^2-s^2)-4sQr+ \frac{1}{4}(4s^2-\kappa)^2 - Q^2 \, , \\
 P &= - \frac{4s^2-\kappa}{2} \, .
\end{split}
\eeq
and for the one-forms defining the metric we still use \eqref{eq:LeftInvariantSphere} for $\Sigma_0 \cong S^2$, but otherwise we locally define
\beq
\begin{aligned}
\Sigma_1 &\cong T^2 &\qquad \sigma_1 + \ii \sigma_2 &= \e^{\ii \tau}(\rd \phi - \ii \, \rd\vartheta) \, , &\qquad \sigma_3 &= \rd\tau - \vartheta \, \rd\phi \, , \\
\Sigma_g \, \ & g >1 &\qquad \sigma_1 + \ii \sigma_2 &= \e^{\ii \tau}(\sinh\vartheta \, \rd \phi - \ii \, \rd\vartheta) \, , &\qquad \sigma_3 &= \rd\tau - \cosh\vartheta \, \rd\phi \, .
\end{aligned}
\eeq
Again, the Killing vector constructed from supersymmetry is $\xi = \partial_\tau$, which generates the fibration and whose norm vanishes at the zeroes of $\Omega$. If $Q = \pm \frac{1}{2}\sqrt{4s^2-\kappa}$, then we can only have $\kappa=1$ and we obtain the Taub-NUT-AdS solution considered earlier, otherwise smoothness of the solution at $r_0$, the largest root of $\Omega$, fixes $Q$ by imposing
\beq
\frac{r_0^2 - s^2}{\abs{\Omega'(r_0)}} = \frac{2s}{\mathbf{p}} \, , \qquad \text{with} \ \mb{p} = \frac{p}{\abs{g-1}} \ \text{for $g \neq 1$, and} \ \mb{p} = p \ \text{for $g=1$.}
\eeq
Analogously, the period of $\tau$ is fixed to be
\beq
\tau \sim \tau + \frac{4\pi}{\mb{p}} \, .
\eeq
There are two branches of solutions, whose existence depends on the intricate relation between the deformation parameter $s$ and the integer $p$. We refer to these branches as ``Bolt$_\pm$'', consistently with \cite{Martelli:2012sz, Toldo:2017qsh}. For both branches the Killing vector $\xi$ has a bolt at $r=r_0$ that is diffeomorphic to $\Sigma_g$, the basis of the fibration. Studying the form of the functions defined in Section  \ref{subsec:LocalForm} near the bolt shows that for the solution Bolt$_\pm$ we have a bolt$_\mp$ in our language: the chirality of the spinor is opposite to the sign identifying the branch of solutions. Thus we can apply \eqref{eq:IOS_IR} and find in the IR
\beq
\label{eq:IBoltIR}
I^{\rm IR}[\text{Bolt}_\pm] = \left( \frac{\pi}{2G_4} \pm \frac{\ii\vartheta}{2} \right) \left( 1-g \mp \frac{p}{4} \right) \, ,
\eeq
since $T\Sigma_g \cong \mc{O}(2-2g)$ and $N\Sigma_g \cong \mc{O}(-p)$.\\ 
The conformal boundary is the manifold $\mc{M}_{g,p}$, a $S^1$ bundle of degree $p$ over the Riemann surface $\Sigma_g$, which includes the $S^3/\Z_p$ bounding the self-dual Taub-NUT-AdS solution. Indeed, the structure of the UV contribution is not very different from \eqref{eq:IUV14BPS}: again $\eta_{(0)} = 2s \, \sigma_3$ and $\rd\eta_{(0)} = 2s \, \vol(\Sigma_g)$. The metric on the Riemann surface is normalized so that $R_{2d}=2\kappa$ and $\vol(\Sigma_g)$ integrates to $\frac{4\pi p}{\mb{p}}$. Therefore \eqref{eq:IOS_UV} gives
\beq
\label{eq:IBoltUV}
I^{\rm UV}= \frac{\ii\vartheta}{2} \frac{ (16 s^4 -8s^2\kappa ) p}{\mb{p}^2} = \begin{cases}
\frac{\ii\vartheta}{2} \frac{ (16 s^4 -8s^2\kappa )(g-1)^2 }{p} \qquad & g \neq 1 \\[5pt]
\frac{\ii\vartheta}{2} \frac{ 16 s^4 }{p} \qquad & g = 1
\end{cases} \, .
\eeq
Notice that this reduces to \eqref{eq:IUV14BPS} for $g=0, \kappa = 1$. Moreover, for the trivial fibration $p=0$ $I^{\rm UV}=0$ and \eqref{eq:IBoltIR} corresponds to \eqref{eq:IOSBH} provided we identify the sign of the branch of the bolt solution with $-\sgn(Q)$.

\subsection{\texorpdfstring{$\frac{1}{2}$}{1/2}-BPS solutions}
\label{subsec:12BPS}

There are also known solutions preserving four real supercharges, such as those constructed in \cite{Martelli:2012sz}. The local form of the metric is again \eqref{eq:MetricGaugeField14BPS}, but now the forms $\sigma_i$ are always the left-invariant one-forms on $S^3$ \eqref{eq:LeftInvariantSphere}.

\medskip

The simplest case is again the Einstein Taub-NUT-AdS solution on $\R^4/\Z_p$ \cite{Martelli:2011fw, Farquet:2014kma}, but now with a different instanton:
\beq
\Omega(r) =  (r-s)^2 \left[ 1 + (r-s)(r+3s) \right] \, , \qquad Q = P = - s \sqrt{4s^2-1} \, .
\eeq
The Killing vector determined by supersymmetry depends on two numbers $(\mathtt{p}, \mathtt{q})\in \C^2\setminus \{ 0\}$. As showed in \cite{BenettiGenolini:2019jdz}, the final result does not depend on the choice of $\mathtt{p}, \mathtt{q}$ provided one of the two vanishes, so we choose $\mathtt{p}=1, \mathtt{q}=0$. Then
\beq
\xi = -2 \left[ (2s+\sqrt{4s^2-1}) \partial_\phi + \left( \frac{1}{2s} - 2s - \sqrt{4s^2-1} \right) \partial_\tau \right]
\eeq
In contrast with the $\frac{1}{4}$-BPS solutions considered earlier, we see that $\xi \neq \partial_\tau$, so the Killing vector induced by supersymmetry is not the one generating the fibration of the solution. Nonetheless, one can show that $\xi$ has a nut at the origin of $\R^4$ at $r=s$: in our language it is a nut$_+$ with 
\beq
b_1 = - \frac{1}{2s} \, , \qquad b_2 = - \frac{1}{2s} +4s + 2 \sqrt{4s^2-1} \, .
\eeq
Thus \eqref{eq:IOS_IR} gives
\beq
I^{\rm IR} = \left( \frac{\pi}{2G_4} - \frac{\ii\vartheta}{2} \right) 4s^2 \, .
\eeq
Despite the fact that the conformal boundary of these solutions is geometrically $S^3/\Z_p \cong \mc{M}_{0,p}$ as for the solutions considered in the previous section, the contribution to the action is different, because the Killing vector $\xi$ is different. This is due to the fact that these solutions, and hence their boundary, preserves four supercharges instead of two. In this case, application of \eqref{eq:IOS_UV} leads to
\beq
I^{\rm UV} = \frac{\ii\vartheta}{2}\frac{16s^4}{p} \, ,
\eeq
and so
\beq
\label{eq:IUV12BPS}
I_{\vartheta} = \frac{\ii\vartheta}{2} 4s^2 \left( \frac{4s^2}{p}-1 \right) \, , 
\eeq
which is consistent with \cite{Martelli:2012sz}.

\medskip

As for the $\frac{1}{4}$-BPS solutions, there is a family of non-self-dual solutions with topology $\mc{O}(-p)\to S^2$ with 
\beq
\begin{split}
\Omega &= (r^2-s^2)^2+ (1 - 4s^2)(r^2-s^2)-2Q\sqrt{4s^2-1} r+ s^2(4s^2-1) - Q^2 \, , \\
 P &= - s\sqrt{4s^2-1} \, .
\end{split}
\eeq
The parameter $Q$ is fixed by regularity of the metric at $r_0$, the largest root of $\Omega$, and this condition identifies two branches of solutions. Since the form of metric is again \eqref{eq:MetricGaugeField14BPS}, the locus $\{ r = r_0 \}$ is a $S^2$ bolt for the Killing vector $\partial_\tau$ for the $\frac{1}{2}$-BPS solutions as well, therefore in the literature the two branches of solutions are again called ``Bolt$_\pm$''. The difference with the previous case is that now $\xi \neq \partial_\tau$, and $\xi$ doesn't have a bolt. In fact, $\xi$ vanishes only at the poles of the $S^2$. More precisely, for the Bolt$_\pm$ solution, the north pole is a nut$_\pm$ with weights
\beq
b_1 = - 4s - 2 \sqrt{4s^2-1} \, , \qquad b_2 = \frac{p}{2s} \, ,
\eeq
and the south pole is a nut$_\mp$ with weights
\beq
b_1 = 4s + 2 \sqrt{4s^2-1} \, , \qquad b_2 = \frac{p}{2} \left( \frac{1}{s} - 8s -4 \sqrt{4s^2-1} \right) \, .
\eeq
Applying \eqref{eq:IOS_IR} then gives
\beq
\label{eq:IBolt12IR}
I^{\rm IR}[\text{Bolt}_\pm] = \frac{\pi}{2G_4} \left[ 1 \pm \frac{2\sqrt{4s^2-1}}{sp} \left( s^2 - \frac{p^2}{16} \right) \right] + \frac{\ii\vartheta}{2} \left( - \frac{p}{4} - \frac{4s^2}{p} \right)
\eeq
Since the conformal boundary data are the same for self-dual and non-self-dual solutions, the contribution from the conformal boundary is again \eqref{eq:IUV12BPS}.

\subsection{Other topologies}
\label{subsec:OtherTopologies}

For the examples considered so far we know an analytic expression for the metric and gauge field. However, this is generically hard to find. The formulae found here and in \cite{BenettiGenolini:2019jdz} only require knowledge of topological data at the fixed points of the $U(1)$ action, without the details of the fields in the bulk. 

For instance, suppose that $\bulk =\cO(-p)\to S^2$, assume the existence of a supersymmetric solution on $\bulk$, and that the isometry generated by $\xi$ is contained in the natural $U(1)^2$ acting on fiber and base. Then one can use simple toric geometry to find an expression for the on-shell action contribution \eqref{eq:IOS_IR} by straightforward generalization of \cite{BenettiGenolini:2019jdz}. In particular, a solution will have a Killing vector $\xi$ with components $(a_1, a_2)$ on the basis of the torus action, and this Killing vector will have two nuts with chiralities $\kappa_1, \kappa_2$ and weights $b_i^{(1)}, b_i^{(2)}$. The IR contribution to the on-shell action is
\begin{align}
\label{eq:IIROpS2}
I^{\rm IR} &= \sum_{i=1}^2 -\kappa_i \left( \frac{\pi}{2G_4} - \kappa_i \frac{\ii\vartheta}{2} \right) \frac{\left( b_1^{(i)} - \kappa_i b_2^{(i)} \right)^2}{4 b_1^{(i)}b_2^{(i)}} \\
\nonumber
&= \frac{1}{4a_1(a_1+a_2)(a_1(p-1)-a_2)}\Big( \mc{Q}_p(\kappa_1, \kappa_2; a_1, a_2) \frac{\pi}{2G_4} - \mc{R}_p(\kappa_1, \kappa_2; a_1, a_2) \frac{\ii\vartheta}{2} \Big)
\end{align}
where
\beq
\begin{split}
\mc{Q}_p(\kappa_1,\kappa_2; a_1,a_2)  &= a_1^3 \left[\kappa_1 \left(p^2-2 p+2\right)+2 (\kappa_2+2) (p-1)\right]+a_2^3 (\kappa_1-\kappa_2) \\ 
& \ \ \  + a_1^2 a_2 (p-2) [\kappa_1 (p-2)  +2 (\kappa_2+2)] \\
& \ \ \  + a_1 a_2^2 [\kappa_1 (3-2 p)+\kappa_2 (p-3)-4] \, , \\[5pt]
\mc{R}_p(\kappa_1,\kappa_2; a_1,a_2)  &= a_1^3 \left[ p^2+2 (p-1) (\kappa_1+\kappa_2)\right] + a_1^2 a_2 (p-2) \left[ 2 (\kappa_1+\kappa_2)+p \right] \\
& \ \ \ - a_1 a_2^2 \left[ 2 (\kappa_1+\kappa_2)+p \right] \, .
\end{split}
\eeq
Using this formula we recover $I^{\rm IR}[{\rm Bolt}_\pm]$ in \eqref{eq:IBoltIR} with $\kappa_1 = \kappa_2 = \mp 1$ and $a_1 = 0$. Similarly, we recover $I^{\rm IR}[{\rm Bolt}_\pm]$ in \eqref{eq:IBolt12IR} when substituting $\kappa_1 = -\kappa_2 = \mp 1$ and
\beq
a_1 = -2 \left( 2s+\sqrt{4s^2-1} \right) \, , \qquad a_2 = -2 \left( \frac{p}{4s} -2s - \sqrt{4s^2-1} \right) \, .
\eeq

The conformal boundary of $\bulk$ is a squashed Lens space $L(p,1) \cong S^3/\Z_p$, with fibration determined by the choice of Killing vector. Thus, the concrete expression for \eqref{eq:IOS_UV} will depend on $a_1, a_2$. However, we can also consider more general vector spaces $L(p,q)$, which are toric $3$-manifolds. Thus, the toric geometry techniques used earlier for $\mc{O}(-p) \to S^2$ can be applied to this case as well, assuming that the torus action extends to the bulk. The case considered in \cite{BenettiGenolini:2019jdz} is that of $L(3,2)$, for which we expect three nuts with equal chiralities $\kappa$:
\beq
\begin{split}
I^{\rm IR}_{L(3,2)} &= \sum_{i=1}^3 -\kappa_i \left( \frac{\pi}{2G_4} - \kappa_i \frac{\ii\vartheta}{2} \right) \frac{\left( b_1^{(i)} - \kappa_i b_2^{(i)} \right)^2}{4 b_1^{(i)}b_2^{(i)}} \\
&= \left( \kappa \frac{\pi}{2G_4} - \frac{\ii \vartheta}{2} \right) \frac{3 \left( 2 a_1(a_1-a_2)(1+\kappa) -a_2^2(3+4\kappa) \right) }{4 (a_1-2a_2)(a_1+a_2)}
\end{split}
\eeq
One of the interesting features of this is example is that $L(3,2)$ is homeomorphic to $L(3,1)$. The action of the filling of the latter can be computed using \eqref{eq:IIROpS2}, obtaining a different result and showing the effects of different supersymmetric structures even for topologically equivalent backgrounds
\beq
I^{\rm IR}_{L(3,1)} = \left( \kappa \frac{\pi}{2G_4} - \frac{\ii \vartheta}{2} \right) \frac{3 \left( a_1^2 (9+8\kappa) + a_1a_2(3+4\kappa) -a_2^2(3+4\kappa) \right) }{4 (2a_1-a_2)(a_1+a_2)} \, .
\eeq

Given any four-manifold with torus action, we can also investigate the behaviour of the on-shell action and the $\vartheta$ term under blow-up, corresponding to substituting a point in $\R^4$ with $\mc{O}(-1)\to S^2$. Using the toric geometry description in \cite{BenettiGenolini:2019jdz}, it is easy to see that if the original nut had chirality $\kappa$ and the solution had on-shell action $I$, then the action of the solution with blown-up topology is
\beq
I_{\text{blow up}} = I + \frac{2+3\kappa}{4} \left( \frac{\pi}{2G_4} - \kappa \frac{\ii\vartheta}{2} \right) \, .
\eeq
Indeed, the $\frac{1}{4}$-BPS solution Bolt$_+$ with topology $\mc{O}(-1)\to S^2$ can be viewed as the blow up of the $\frac{1}{4}$-BPS Taub-NUT-AdS solution (the former has a bolt$_-$ and the latter a nut$_-$), and the results for the on-shell action match.

\section{Origin of the \texorpdfstring{$\vartheta$}{theta} term}
\label{sec:OriginTheta}

Four-dimensional minimal supergravity arises from the consistent truncation of ten- and eleven-dimensional supergravities on numerous six- and seven-dimensional internal spaces (potentially with fluxes). In the context of top-down constructions of the AdS/CFT correspondence, different choices of internal manifolds and fluxes correspond to different field theories, so the four-dimensional minimal supergravity describes the ``universal'' part of the dual three-dimensional $\mc{N}=2$ SCFT corresponding to each internal manifold. 

Here we focus on three-dimensional SCFTs constructed from stacks of $M$-branes in $M$-theory. The corresponding dual geometries are eleven-dimensional space-times with an AdS$_4$ factors, and have been analysed in some generality (see e.g. \cite{Gabella:2012rc}). Each of these geometries should correspond to the uplift of the vacuum solution of an effective four-dimensional gauged supergravity obtained by consistently truncate the eleven-dimensional supergravity on the seven-dimensional internal space \cite{Gauntlett:2007ma}.

We only consider the truncation of the bosonic sector of eleven-dimensional supergravity, which contains the eleven-dimensional metric $g(Y_{11})$ and a three-form gauge field $C$, with curvature $G = \rd C$, interacting via
\beq
\label{eq:11dSUGRA}
S_{11} = \frac{1}{2\kappa^2_{11}}\int \left( R(Y_{11}) \, \vol({Y_{11}}) - \frac{1}{2}G \wedge *_{11}G - \balpha \frac{1}{6}C \wedge G \wedge G \right) \, ,
\eeq
and $2\kappa_{11}^2 = (2\pi)^8 \ell_P^9$. For the purposes of this section we move back to Lorentzian signature and reinstate the gauge couplings. The sign of $\balpha = \pm 1$ is determined by the supersymmetry conventions.

\medskip

As mentioned above, it is possible to obtain four-dimensional minimal gauged supergravity also via consistent truncations of ten-dimensional supergravity. An intriguing possibility is the compactification from massive IIA, which itself breaks parity. In this case, the internal space is a smooth geometry over a space that is topologically $S^6$, constructed as a $S^2$ bundle over $\mathbb{CP}^2$ \cite{Guarino:2015jca, Varela:2019vyd}. The dual field theories are SCFTs obtained from $D2$-branes. Note that the smoothness condition can be relaxed, obtaining internal spaces that consist of singular geometries constructed by replacing the $\mathbb{CP}^2$ base of the bundle with another K\"ahler--Einstein space \cite{Fluder:2015eoa}.

\subsection{\texorpdfstring{$M2$}{M2}-branes}
\label{subsec:M2}

For the case of field theories obtained from arrangements of $M2$ branes, there are numerous consistent truncations known. Equivalently, it is known that for any supersymmetric solution to the equations of motion of minimal gauged supergravity \eqref{eq:4dEOMs} there are numerous ways of uplifting on a seven-dimensional manifold to obtain an eleven-dimensional configuration for which the integral of $*_{11}G$ through a seven-cycle is non-zero, see e.g. \cite{Duff:1984hn, Gauntlett:2007ma, Larios:2019lxq}. Here we consider one such ansatz. Inspired by the Freund--Rubin AdS$_4\times S^7$ solution, we take the ansatz $\bulk \times SE_7$, where $\bulk$ is the four-dimensional spacetime and $SE_7$ is a seven-dimensional Sasaki--Einstein manifold  \cite{Gauntlett:2007ma}.\footnote{There are restrictions on the global structure of the $SE_7$, depending on the topology and spin$^c$ structure on $\bulk$ \cite{Martelli:2012sz, Toldo:2017qsh}. The supersymmetry conventions here are such that $\balpha = +1$.} Locally, we may write the eleven-dimensional metric as
\beq
\label{eq:M2Ansatz}
\begin{split}
g(Y_{11})  &=  \frac{L^2}{\g^2} \left[ \frac{\g^2}{4} g(\bulk) + \left( \rd\psi + \sigma + \frac{\g}{2} A \right)^2 + g(N_6) \right]  \, , \\
G &= \frac{L^3}{\g^3} \left[ \frac{3\g^4}{8} \, \vol(\bulk) - \frac{\g}{2} *_{4} F \wedge J \right]  \, ,
\end{split}
\eeq
where the second factor is a fibering over the local form of the metric on $SE_7$: $\partial_\psi$ is the Reeb vector, $J=\rd\sigma/2$ is the K\"ahler form on the local K\"ahler--Einstein space $N_6$ over which $SE_7$ is constructed. $A$ is a $U(1)$ gauge connection with curvature $F = \rd A$. The metric on $SE_7$ is normalized so that $\Ric(g(SE_7)) = 6 g(SE_7)$.

This ansatz gives a consistent truncation of eleven-dimensional supergravity down to minimal four-dimensional supergravity, in the sense that upon substitution in the equations of motion obtained from the action \eqref{eq:11dSUGRA}, we obtain the equations of motion \eqref{eq:4dEOMs}. Once this has been checked, we may insert the ansatz back in the action $S_{11}$: could there be a $\vartheta$ term from the eleven-dimensional topological term? We have
\beq
G \wedge G = - \frac{L^6}{4\g^4} J \wedge J \wedge F \wedge F \, ,
\eeq
so to find a $\vartheta$ term in four dimensions we would need a term $\Psi_3$ in $C$ entirely along $SE_7$, which is only compatible with the ansatz \eqref{eq:M2Ansatz} if such term is closed. In order to have a non-trivial integral on $SE_7$ when reducing, we need $\partial_\psi \hook \Psi_3 \neq 0$. However, it is easy to see that this requirement cannot be combined with closedness, in the sense that together they imply that $\Psi_3 \wedge G \wedge G = 0$. In fact, this argument can be extended to different Freund--Rubin-type reductions: it is not possible to include a non-trivial pullback $G|_{SE_7}$, and thus have a non-trivial $C|_{SE_7}$ \cite{Benishti:2010jn}. Therefore, we conclude that generically \textit{there is no $\vartheta$ term} in the supergravity action obtained by reduction of eleven-dimensional supergravity on $SE_7$, which is holographically dual to three-dimensional SCFTs with $\mc{N}=2$ obtained from $M2$-branes.

In order to compare the field theory and gravity observables, the uplift \eqref{eq:M2Ansatz} gives the relation between the four-dimensional Newton constant and the quantization of the fluxes for $G$. Consistently with the interpretation as gravity dual to $M2$ branes, $G$ has an electric flux proportional to $\vol(\bulk)$, and
\beq
N_{M2} = - \frac{1}{(2\pi \ell_P)^6}\int_{SE_7} *_{11}G =  \frac{6 L^6 {\rm Vol}(SE_7)}{\g^6(2\pi\ell_P)^6}
\eeq
Combining with the reduction of the Ricci scalar, we find
\beq
\label{eq:G4M2}
\frac{1}{16\pi G_4} = \frac{\pi \, \g^2}{12 \sqrt{6 {\rm Vol}(SE_7)}}  N_{M2}^{3/2}  \, .
\eeq
This indeed reproduces the large $N$ behaviour $N^{3/2}$ typical of $M2$ branes.

\subsection{\texorpdfstring{$M5$}{M5}-branes}
\label{subsec:M5}

The other maximally symmetric Freund--Rubin solution AdS$_7\times S^4$ is dual to the worldvolume theory of a stack of $M5$-branes. The corresponding consistent truncation was found in \cite{Nastase:1999cb, Nastase:1999kf}.\footnote{Given the importance of the gauge fields in the following, we should point out that the consistent truncation from eleven to seven dimensions has only been worked out in full with the gauge fields set to zero. More precisely, the reduction of the eleven-dimensional Einstein equations of motion has been done with vanishing gauge fields. Nonetheless, there is substantial supporting evidence in favour of the consistency, including the full reduction of the supersymmetry variations \cite{Nastase:1999kf}, the reduction of less supersymmetric supergravities \cite{Lu:1999bc}, and various limits of the ansatz, such as \cite{Cvetic:2000ah}. In this case, we should pick $\balpha=-1$ in \eqref{eq:11dSUGRA} to be consistent with the supersymmetry conventions.} Using the notations of \cite{Cvetic:2000ah}, we write the eleven-dimensional metric as
\beq
\label{eq:g11}
g(Y_{11}) = \frac{\Delta^{1/3}}{\g^2} \left[ \g^2 g(Y_7) + \Delta^{-1}T_{ij}^{-1} D\mu^i D\mu^j \right]
\eeq
and the four-form as
\beq
\label{eq:G4ConsistentUplift}
\begin{split}
G &= \frac{\Delta^{-2}}{\g^3 4!} \epsilon_{i_1\dots i_5} \bigg( -U \, \mu^{i_1}D\mu^{i_2} \wedge D\mu^{i_3} \wedge D\mu^{i_4} \wedge D\mu^{i_5} \\
& \qquad \qquad \quad \quad \ + 4 \, T^{i_1m} \, DT^{i_2n} \wedge \mu^m \, \mu^n \, D\mu^{i_3} \wedge D\mu^{i_4} \wedge D\mu^{i_5} \\
&  \qquad \qquad \quad \quad \ + 6 \g \, \Delta \, \cF^{i_1i_2} \wedge D\mu^{i_3} \wedge D\mu^{i_4} T^{i_5j}\mu^j \bigg) \\
& \ \ \  - T_{ij} *_7 S^i_{(3)} \mu^j + \frac{1}{\g} S^i_{(3)} \wedge D\mu^i
\end{split}
\eeq
In these expressions, $\g$ is the gauge coupling, $i,j=1, \dots 5$, and we have introduced the following objects: coordinates $\mu^i$ on $S^4$ satisfying $\sum \mu^i \mu^i = 1$; three-forms $S_{(3)}^i$; $SO(5)$ gauge fields $\cA^{ij}_{(1)}$ with curvature $\cF_{(2)}^{ij}$; and fourteen scalar fields grouped in the symmetric unimodular matrix $T^{ij}$. Moreover, we have defined
\beq
\begin{aligned}
\Delta &\equiv T_{ij}\mu^i \mu^j \, , &\qquad U &\equiv 2T_{ij}T_{jk}\mu^i\mu^k - \Delta \, T_{ii} \, , \\
D\mu^i &= \rd\mu^i + \g \cA^{ij}\mu^j \, , & \qquad
D T_{ij} &= \rd T_{ij} + \g \cA^{ik}T_{kj} + \g \cA^{jk} T_{ki} \, , \\
D S^i_{(3)} &= \rd S^i_{(3)} + \g \cA^{ij} \wedge S^j_{(3)} \, , &\qquad \cF^{ij} &= \rd \cA^{ij} + \g \cA^{ik} \wedge \cA^{kj} \, .
\end{aligned}
\eeq

Inserting this ansatz into the equations of motion and action of eleven-dimensional supergravity takes us to the $SO(5)$ gauged maximal supergravity in seven dimensions \cite{Pernici:1984xx} 
\beq 
\label{eq:7dSUGRAAction}
\begin{split}
S_7 &= \frac{1}{2\kappa^2_7}\int \Big[ (R-V) \, \vol (Y_7) - \tfrac{1}{4} T_{ij}^{-1} DT_{jk} \wedge *_7 T^{-1}_{kl} DT_{li} - \tfrac{1}{4} T^{-1}_{ik}T^{-1}_{jl} \cF^{ij} \wedge *_7 \cF^{kl} \\
& \qquad \qquad \ \ \  - \tfrac{1}{2} T_{ij} S_{(3)}^{i} \wedge *_7 S_{(3)}^j + \tfrac{1}{2\g} S^i_{(3)}\wedge DS^i_{(3)} - \tfrac{1}{8\g} \epsilon_{ij_1\dots j_4} S^i_{(3)} \wedge \cF^{j_1j_2} \wedge \cF^{j_3j_4}  \\
& \qquad \qquad \ \ \ + \tfrac{1}{16\g} Q_3[\cA, \cF] \wedge P_4[\cF]  - \tfrac{1}{8\g} Q_7[\cA,\cF]   \Big] \, ,
\end{split}
\eeq
where the scalar potential is
\beq
V = \tfrac{1}{2}\g^2 \left( 2 T_{ij}T_{ij} - (T_{ii})^2 \right) \, ,
\eeq
and the seven-dimensional Chern--Simons forms are given by
\begin{gather}
\label{eq:Omega3}
Q_3[A,F] \wedge P_4[F] = \tr \left( A \wedge F - \frac{1}{3}\g \, A \wedge A \wedge A \right) \wedge \tr \left( F \wedge F \right) \, , \\[5pt]
\label{eq:Omega5}
\begin{split}
Q_7[A,F] &= \tr \Big( A \wedge F \wedge F - \frac{2}{5}\g \, A \wedge A \wedge A \wedge F \wedge F - \frac{1}{5}\g \, A \wedge F \wedge A \wedge A \wedge F \\
&\qquad \ \  + \frac{1}{5}\g^2 \, A \wedge A \wedge A \wedge A \wedge A \wedge F \\
& \qquad - \frac{1}{35} \g^3 \, A \wedge A \wedge A \wedge A \wedge A \wedge A \wedge A \Big) \, .
\end{split}
\end{gather}
Solutions to this supergravity uplift to eleven-dimensional solutions, and describe holographic duals to the six-dimensional $(2,0)$ SCFT on spaces diffeomorphic to their conformal boundary. The relation between the eleven-dimensional and seven-dimensional Newton constants is found by looking at the reduction of the action:
\beq
\label{eq:kappa7kappa11}
\frac{1}{2\kappa^2_7} = \frac{\Vol(S^4)}{(2\pi)^8\ell_P^9\g^4} \, .
\eeq
Compared with the previous ansatz, we immediately see that $G$ satisfies a different flux quantisation condition, corresponding to the presence of $M5$ branes\footnote{Generically, it is not $G$ that has quantised flux, but rather \cite{Witten:1996md}
\[
\left[ \frac{1}{(2\pi\ell_P)^3}G \right] - \frac{1}{4}p_1(Y_{11}) \in H^4(Y_{11},\Z) \, .
\]
However, the Pontryagin class of any spin manifold of dimension equal or lower than seven is a multiple of $4$, so the quantization condition is not corrected for our purposes.}
\beq
N_{M5} = \frac{1}{(2\pi\ell_P)^3}\int_{S^4} G = \frac{1}{(2\pi\ell_P)^3}\int_{S^4} \left( - \frac{\Delta^{-2}U}{\g^3} \vol(S^4) \right) \, .
\eeq
Combining the two, we find
\beq
\label{eq:2kappa7}
\frac{1}{2\kappa^2_7} = 2\pi \g^5\frac{\Vol(S^4)}{3\left(- \int_{S^4} \Delta^{-2}U \, \vol(S^4) \right)^3 }  N_{M5}^3 
\eeq
For instance, the vacuum solution has metric $g(Y_7) = 4 g(AdS_7)/\g^2$, where $g(AdS_7)$ is normalised to have sectional curvature equal to $-1$, and non-vanishing $T_{ij} = \delta_{ij}$. It uplifts to eleven dimensions on $S^4$ as the Freund--Rubin solution, that is, without fibering and warping, and
\beq
\frac{1}{2\kappa^2_7} = \frac{\g^5}{96\pi^3} N_{M5}^3  \, .
\eeq

\medskip

Our interest, though, is in supersymmetric solutions containing a three-dimensional hyperbolic factor, such as that found in \cite{Pernici:1984nw}\footnote{There is also a non-supersymmetric vacuum anti-de Sitter solution, which is not relevant for us.}
\beq
\label{eq:PerniciSezgin}
\begin{split}
g(Y_7) &= \frac{2^{\frac{4}{5}}}{\g^2} \left[ g({\rm AdS}_4) + {g}(\internal) \right] \, , \qquad T = {\rm diag} \left( 2^{- \frac{2}{5}}, 2^{- \frac{2}{5}}, 2^{- \frac{2}{5}}, 2^{\frac{3}{5}}, 2^{\frac{3}{5}} \right) \\ 
\cA^{ab} &= \tfrac{1}{\g}{\omega}^{ab} \, ,
\end{split}
\eeq
where $\internal$ is a closed hyperbolic $3$-manifold, and both $g({\rm AdS}_4)$ and ${g}(\internal)$ are normalized to have constant sectional curvature equal to $-1$. We have broken $SO(5)$ to $SO(3)\times SO(2)$, splitting the indices $i,j$ into $a,b = 1, 2, 3$ and $\alpha, \beta = 1, 2$, and set the $SO(3)$ gauge fields proportional to the spin connection ${\omega}$ of $\internal$. This solution describes $M5$ branes wrapped around $\internal$, which is showed by the presence of a gauge field fixed by the geometry, corresponding in the field theory setup to a non-trivial background gauge field realizing the topological twist \cite{Gauntlett:2000ng}.

As we always expect, the presence of this solution signals the existence of a consistent truncation to an effective four-dimensional gauged supergravity with anti-de Sitter vacuum. In fact, there is a consistent truncation to four-dimensional $\cN=2$ gauged supergravity with one vector multiplet and two hypermultiplets \cite{Donos:2010ax}, but here we will only be concerned with a truncation to minimal gauged supergravity. It is obtained by the following ansatz
\beq
\label{eq:M5Ansatz}
\begin{aligned}
g(Y_7) &= \frac{2^{\frac{4}{5}}}{\g^2} \left[ \g^2 g(\bulk) + {g}(\internal) \right] \, , & \qquad T &= {\rm diag} \left( 2^{- \frac{2}{5}} , 2^{- \frac{2}{5}} , 2^{- \frac{2}{5}} , 2^{\frac{3}{5}} , 2^{\frac{3}{5}} \right) \, , \\
\cA^{ab} &= \frac{1}{\g}{\omega}^{ab} \, , \qquad \cA^{\alpha\beta} = 2\epsilon^{\alpha\beta} A \, , &\qquad S_{(3)}^a &= - \frac{2}{\g} *_4 F \wedge {\e}^a  \, ,
\end{aligned}
\eeq
where $A$ is a $U(1)$ gauge field on $\bulk$ with curvature $F=\rd A$, ${\e}^a$ and ${\omega}^{ab}$ are the dreibein and spin connection of the hyperbolic metric on $\internal$.
Inserting this ansatz into the equations of motion derived from \eqref{eq:7dSUGRAAction}, leads to the equations of motion of four-dimensional minimal gauged supergravity \eqref{eq:4dEOMs}. Notice for instance that the vacuum solution $g(AdS_4)/\g^2$ with vanishing gauge field uplifts to the Pernici--Sezgin solution \eqref{eq:PerniciSezgin}.

Having showed consistency, we can then substitute the ansatz back inside the action \eqref{eq:7dSUGRAAction}. Whilst the first two lines give an expression proportional to $\Vol(\internal)$, the topological term in the last line of \eqref{eq:7dSUGRAAction} results in a purely topological term in four dimensions as well. To be more precise, we write a bulk manifold $X_8 \cong X_4 \times Y_4$ with $\partial X_4 \cong \internal$, so clearly $\partial X_8 \cong Y_7$. Then the seven-dimensional topological term may be written as
\beq
\begin{split}
\frac{1}{2\kappa^2_7}\frac{1}{16\g}\int_{X_8} \left( P_4[\cF]^2 - 2 P_8[\cF] \right) &= - \frac{1}{2\kappa^2_7\g^3} \int_{X_4} P_4[\overline{\rho}] \, \int_{Y_4} F \wedge F \\
&= - \frac{8\pi \, cs(\internal)}{2\kappa^2_7\, \g^3} \int_{Y_4} F \wedge F
\end{split}
\eeq
Here, $\overline{\rho}$ is the extension in the bulk $X_4$ of the curvature two-form of $\omega$, and ${cs}(\internal)$ is the three-dimensional Chern--Simons invariant of the three-manifold $\internal$
\beq
\label{eq:CSHyperbolic}
cs(\internal) = \frac{1}{8\pi} \int_{\internal} \tr_{\mb{3}} \left( {\omega} \wedge \rd {\omega} + \frac{2}{3} {\omega} \wedge {\omega} \wedge {\omega} \right) \, ,
\eeq
where the trace is in the defining representation of $\mf{so}(3)$. It is defined modulo $2\pi$ having chosen a spin structure on $\internal$ (see Appendix \ref{sec:CSInvariant}).

The main difference with the $M2$-brane ansatz \eqref{eq:M2Ansatz} is that now we find an additional $\vartheta$ term due to the reduction of the seven-dimensional topological terms
\beq
\label{eq:ReductionS7}
S_7 = \frac{4 \Vol(\internal)}{2\kappa^7 \g^3} \int_{\bulk} \left( R (g(\bulk)) + 6\g^2 - F^2 \right) \vol(\bulk) - \frac{8\pi \, cs(\internal) }{2\kappa^2_7\g^3} \int_{\bulk} F \wedge F  \, .
\eeq

In order to find the relation between the four-dimensional gravitational constants and the M-theory quantities, we need an expression for $\Delta$ and $U$ to substitute in \eqref{eq:2kappa7}. To find that, we plug the ansatz \eqref{eq:M5Ansatz} in the uplift \eqref{eq:g11}. To be more concrete, following \cite{Gang:2014ema, Bobev:2019zmz}, we introduce the following coordinates $\mu^i$ on $S^4$:
\begin{equation}
\begin{split}
\mu^a = \,&\cos(\nu) \, \tilde{\mu}^a\,, \qquad a=1,2,3\,,\\
 \mu^{\alpha}=\,&\sin(\nu) \, \tilde{\mu}^\alpha\,,\qquad \alpha=4,5\,,
\end{split}
\end{equation}
with $\sum_{i=1}^{5}(\mu^{i})^2=\sum_{a=1}^{3}(\tilde{\mu}^a)^2=\sum_{\alpha=4}^{5}(\tilde{\mu}^{\alpha})^2=1$. Furthermore, we use the explicit parametrization 
\beq
\begin{aligned}
\tilde{\mu}^1&=\cos\xi_1\,, &\quad \tilde{\mu}^2&=\sin\xi_1\cos\xi_2\,, &\quad \tilde{\mu}^3&=\sin\xi_1\sin\xi_2\,, \\
\tilde{\mu}^4&=\cos\psi\,, &\qquad  \tilde{\mu}^5&=\sin\psi\, .
\end{aligned}
\end{equation}
with coordinate ranges  $0\leq \nu\leq \pi/2$, $0\leq \psi< \Delta\psi$, $0\leq \xi_{1}\leq \pi$, and $0\leq \xi_{2}<2\pi$. Using these coordinates, we have
\beq
\label{eq:DeltaUAnsatz}
\Delta = 2^{-2/5}(1+\sin^2\nu) \, , \qquad U = - 2^{-4/5}(5+\sin^2\nu) \, .
\eeq
The eleven-dimensional ansatz for the metric is written
\beq
\label{upliftM5}
\begin{split}
g(Y_{11}) = \frac{2^{\frac{2}{3}}(1+\sin^2\nu)^{\frac{1}{3}}}{\g^2} \Big[ & \g^2 g(\bulk) + {g}(\internal) + \frac{1}{2} \Big(\rd\nu^2+ \frac{\sin^2\nu}{1+\sin^2\nu}\left( \rd\psi- \g A \right)^2 \Big) \\
&\qquad+ \frac{\cos^2\nu}{1+\sin^2\nu}\sum_{a=1}^{3}\left( \rd\tilde{\mu}^a + \omega^{ab}\tilde{\mu}^b\right)^2 \Big] \,,
\end{split}
\eeq
and the four-form gauge field is
\begin{equation}
\label{G4M5}
\begin{split}
G =\,& \frac{(5+\sin^{2}\nu)}{\g^3(1+\sin^{2}\nu)^{2}}\,  \epsilon_{abc}\epsilon_{\alpha\beta}\, D\mu^{b}\wedge D\mu^{c}\wedge \left( \frac14 \mu^{a} D\mu^{\alpha}\wedge D\mu^{\beta}+\frac16\mu^{\alpha}D\mu^{\beta}\wedge D\mu^{a} \right) \\
\,&+\frac{\epsilon_{abc}\epsilon_{\alpha\beta} }{\g^3(1+\sin^{2}\nu)} {\rho}^{ab} \wedge \left( D\mu^{c}\wedge D\mu^{\alpha} \mu^{\beta}+\frac{1}{4}  D\mu^{\alpha}\wedge D\mu^{\beta}\mu^{c} \right) \\
& \, +\frac{\epsilon_{abc}}{\g^2(1+\sin^{2}\nu)} F \wedge D\mu^{a}\wedge D\mu^{b}\mu^{c} - \frac{2}{\g^2} \ast_{4}F \wedge \e^{a}\wedge D\mu^{a}  - \frac{2}{\g^2} F \wedge *_3 {\e}^{a} \mu^{a}\,,
\end{split}
\end{equation}
where ${\rho}^{ab} = - {\e}^a \wedge {\e}^b$ is the curvature two-form of the spin connection for $\internal$.

Substituting these values in \eqref{eq:2kappa7} and in turn in \eqref{eq:ReductionS7} leads to 
\beq
\label{eq:4dActionLorentzian}
S = \frac{1}{16\pi G_4}\int \left( R_g + 6\g^2 - F^2 \right) \vol_g - \frac{\vartheta\g^2}{8\pi^2}\int F \wedge F
\eeq
with
\beq
\label{eq:UpliftM5}
\frac{1}{G_4} = \frac{8\Vol(\internal) \g^2 }{3\Delta\psi^2} N_{M5}^3 \, , \qquad \vartheta = \frac{8\pi^2 \, {cs}(\internal)}{3\Delta\psi^2}  N_{M5}^3 \, .
\eeq
This reproduces the large $N$ behaviour typical of $M5$-branes. 

\subsection{Uplift: Topological considerations}

The eleven-dimensional solution is constructed as a fibration of $S^4$ over $Y_4$ and $\internal$. The choice of $\internal$ corresponds to the choice of field theory. A necessary requirement is that the full solution must be spin.

If we choose the eleven-dimensional solution to be globally a fibration with $\Sigma_3$ and $S^4$ --- which are spin --- then $Y_4$ is also required to be spin and so $A$ must be a connection on a $U(1)$ bundle. This is fine for many solutions, including AdS$_4$ (even with the non-trivial instanton of \cite{Farquet:2014kma, Martelli:2011fu}); the self-dual Taub-NUT-AdS solutions of \cite{Martelli:2012sz}, which have topology $\R^4$; the SUSY dyonic black holes with topology $D^2\times \Sigma_g$.

However, generically there are also four-dimensional solutions $Y_4$ that are only spin$^c$. For instance, this is the case for the non-self-dual bolt solutions of \cite{Martelli:2012sz}. They have topology $\mc{M}_p \equiv \mc{O}(-p) \to \Sigma_g$ and the $U(1)$ connection has flux through the $\Sigma_g$ cycle. For the $\frac{1}{4}$-BPS, we have
\beq
\g\int_{\Sigma_g}\frac{F}{2\pi} = \pm \frac{p}{2} - (1-g) \, .
\eeq
Since this is a half-integer for odd $p$, $A$ is a spin$^c$ field in that case. 

Looking at the form of the metric in \eqref{upliftM5}, we see that the $S^4$ is written as a $S^1\times S^2$ fibration over an interval: at $\nu=0$ the circle collapses and the $S^2$ collapses at $\nu=\pi/2$. In order to guarantee the regularity of the metric at $\nu=0$ it is important that the global angular form of the circle bundle has the right periodicity. The canonical periodicity of $\psi$ in order to cover the full $S^4$ is $\Delta\psi = 2\pi$. To have a well-defined circle bundle its first Chern number should be an integer, which would also require $A$ to be a well-defined $U(1)$ connection, which doesn't happen for generic $Y_4 = \mc{M}_p$. To remedy this and guarantee that we still have a well-defined circle bundle, we should change the periodicity of $\psi$: by setting $\Delta\psi = 2\pi/k$, the first Chern number of the bundle over $\Sigma_g$ is
\beq
\begin{split}
k\g\int_{\Sigma_g}\frac{F}{2\pi} \in \Z \quad &\Leftrightarrow \quad k \left( \pm \frac{p}{2} - (1-g) \right) \in \Z \\
&\Leftrightarrow \quad \pm kp - 2k(1-g) = 0 \mod 2 \, .
\end{split}
\eeq
This requires $kp\in 2\Z$: if $p$ is odd, we need to take $k=2$, for instance. In fact, this is more general than this specific solution: if $A$ is a spin$^c$ connection, we require $\psi$ to have $\pi$ periodicity, resulting in uplifting $Y_4$ on $\Sigma_3\times S^4/\Z_2$.

Having considered the global regularity of the metric, we should then ask whether the eleven-dimensional space $Y_{11}$ is spin. As mentioned, $Y_{11}$ is a sphere bundle of a $\R^5$ bundle over $Y_4\times \internal$, with total space $Z_{12}$. In fact, the way $S^4$ is written in \eqref{upliftM5} shows that it is easier to write $\R^5\cong \R^2\times \R^3$, where $\R^2$ is fibered over $Y_4$ and $\R^3$ is fibered over $\internal$. Since $Y_{11}\cong \partial Z_{12}$, we may focus on $Z_{12}$. More is true: any vector bundle deformation retracts to its zero section, which is isomorphic to the base, so we can just study $Y_4\times \internal$. This is because a deformation retraction is an homotopy equivalence, so $H^\bullet(Z_{12}) \cong H^\bullet(Y_4\times \internal)$. 

The bundle we should focus on is then $\cL\oplus T \internal$, where $\cL$ is the $\R^2$ bundle over $Y_4$. It is tempting to conclude that $Z_{12}$ is spin iff $\cL$ has vanishing Stiefel--Whitney class, but there is a subtlety that we have so far glossed over. As we have found earlier, in order to guarantee regularity of the metric it is necessary to uplift spin$^c$ manifolds on $S^4/\Z_2$, which means that in those cases $\cL$ is a bundle with fiber $\R^2/\Z_2$. Studying its cohomology is subtler, and we shall not attempt it here. We should remark, though, that asking for spin $Z_{12}$ is only necessary in order to guarantee that $Y_{11}$ is spin, so it could be possible to find the uplift conditions in another way. In the following we shall restrict our considerations to spin $\bulk$, which uplift on $S^4$. 

\subsection{Subleading corrections}
\label{subsec:subleading}

Starting from eleven-dimensional supergravity, we can also begin studying the subleading corrections to the large $N$ result. As before, we first reduce to seven-dimensional supergravity, and then to four-dimensional minimal supergravity using the Pernici--Sezgin ansatz. However, in order to simplify our computations we shall focus only on the $SO(5)$ gauge fields, partly following \cite{Harvey:1998bx}. As mentioned in the previous section, we view spacetime $Y_{11}$ as a sphere bundle over $Y_7$, and the resulting seven-dimensional gauge group arising from the structure group of this bundle. The ansatz for the four-form \eqref{eq:G4ConsistentUplift}, which has non-trivial flux through the four-sphere, is an extension of the global angular four-form, which is the unique closed and gauge-invariant extension of the sphere volume form (see Appendix \ref{app:AngularForms}). In fact, we may single out the part of the ansatz with flux through $S^4$ by writing\footnote{As a check, notice that this expression for $G$ gives, using \eqref{eq:E4}
\[
G = \frac{1}{8\pi^2} \epsilon_{a_1a_2a_3a_4a_5} \, y^{a_1} \, D y^{a_2}\wedge D y^{a_3} \wedge D y^{a_4} \wedge D y^{a_5} + \dots
\]
Setting $T_{ij}=\delta_{ij}$ in the full ansatz \eqref{eq:G4ConsistentUplift} gives $\Delta = 1, U=-3$ and so we find the same expression at leading order.}
\beq
G = \frac{8\pi^2}{\g^3} \, \E_4 + \dots \, .
\eeq
A more precise definition of the topological term in the action of eleven-dimensional supergravity \eqref{eq:11dSUGRA} can be given in terms of the integral of a closed form on a twelve-dimensional manifold $X_{12}$ for which $\partial X_{12}= Y_{11}$, which we view as a $S^4$ fibration over $X_8$ (with $\partial X_8 \cong Y_7$). Then, we can use the Bott--Cattaneo formula \eqref{eq:BottCattaneo}
\beq
\label{eq:Naive11dTo7d}
\begin{split}
S_{11} &\supset - \balpha\frac{1}{2\kappa^2_{11}}\frac{1}{6} \int_{X_{12}} G \wedge G \wedge G = - \balpha\frac{1}{2\kappa^2_{11}}\frac{256\pi^6}{3\g^9} \int_{X_{12}} \E_4 \wedge \E_4 \wedge \E_4 \\
&= - \balpha\frac{1}{2\kappa^2_{11}}\frac{64\pi^6}{3\g^5} \int_{X_{8}} p_2(\overline{E}) \\
&= - \balpha\frac{1}{2\kappa^2_{7}} \frac{1}{16\g} \int_{Y_7} \left( Q_3[\cA, \cF] \wedge P_4[\cF] - 2Q_7[\cA, \cF] \right) \, . 
\end{split}
\eeq
Here, $E$ is the rank-5 bundle over $Y_7$ used in the construction of the sphere bundle, and $\cA, \cF$ are, respectively, the connection and curvature of the $SO(5)$ gauge bundle, whereas $\overline{E}$ and $\overline{\cA}$, $\overline{\cF}$ are the corresponding extensions in the bulk $X_8$. In the last equation, we have used \eqref{eq:kappa7kappa11}, and we recognise the topological term in the seven-dimensional supergravity Lagrangian \eqref{eq:7dSUGRAAction}. The choice of $\balpha = -1$, different from the case considered in Section \ref{subsec:M2} is due to the different supersymmetry conventions used also in \cite{Cvetic:2000ah, Nastase:1999kf}.

\medskip

It is known that the lowest-order correction to the equation of motion of the eleven-dimensional four-form is \cite{Vafa:1995fj, Duff:1995wd}
\beq
\label{eq:S1loop}
S_{\rm 1-loop} = - \balpha\frac{2\pi}{(2\pi\ell_P)^3}\int_{X_{12}} G \wedge \frac{p_1(X_{12})^2 - 4p_2(X_{12})}{192} \, .
\eeq
In order to reduce this term to seven dimensions, we expand the Kaluza--Klein ansatz \eqref{eq:g11} to leading order in the derivatives, so that $TX_{12}|_{X_8}\cong TX_8 \oplus E$ and the spin connection is simply the sum of the two connections:
\beq
\omega(X_{12})^{AB} = \begin{pmatrix}
\omega(X_8)^{ab} & 0 \\ 0 & \g \overline{\cA}^{ij} \nabla^m K^{ij,n}
\end{pmatrix} \, ,
\eeq
where $A, B$ are indices on $TX_{12}$, $a,b$ indices on $TX_8$, $\overline{\cA}^{ij}$ are the components of the $SO(5)$ gauge connection, $K^{ij}$ are Killing vectors generating the $SO(5)$ isometries on $S^4$, and $m,n$ are indices on $TS^4$. Therefore, we write
\beq
\label{eq:SplittingPontryagin}
\begin{split}
p_1(X_{12}) &= p_1(X_8) + \g^2 \, p_1(\overline{E}) \, , \\ 
p_2(X_{12}) &= p_2(X_8) + \g^2 \, p_1(X_8)\wedge p_1(\overline{E}) + \g^4 \, p_2(\overline{E}) \, .
\end{split}
\eeq
We can then integrate along the $S^4$ fiber, obtaining the one-loop correction to seven-dimensional supergravity
\beq
\begin{split}
S_{\rm 1-loop} &= - \balpha \frac{1}{24\g^3\ell_P^3} \int_{X_8} \left[ - p_2(X_8) - \g^4 p_2(\overline{E}) + \frac{1}{4}\left( \g^2 p_1(\overline{E}) - p_1(X_8) \right)^2 \right] \, .
\end{split}
\eeq
Altogether, using the quantization of the four-form flux and $\balpha=-1$, the topological terms in seven dimensions read
\beq
\begin{split}
S_{7,{\rm top}} &= 2\pi \int_{X_8} \Bigg[ \frac{N^3-N}{24} \g^4 p_2(\overline{E}) \\
& \qquad \qquad \quad + \frac{N}{48}\left( \g^4 p_2(\overline{E}) - p_2(X_8) + \frac{1}{4}\left( \g^2 p_1(\overline{E}) - p_1(X_8) \right)^2 \right) \Bigg] \, .
\end{split}
\eeq
As expected by AdS/CFT, this reproduces the structure of the $R$-symmetry anomaly of the six-dimensional $(2,0)$ SCFT obtained on the worldvolume of $N_{M5}$ $M5$-branes, which reduces for $N_{M5}=1$ to the anomaly of a free six-dimensional tensor multiplet (the second line).

Now consider the Pernici--Sezgin ansatz \eqref{eq:M5Ansatz}. We have
\beq
\begin{aligned}
p_1(X_8) &= p_1(Y_4) + p_1(X_4) \, , &\qquad p_2(X_8) &= p_1(Y_4) \wedge p_1(X_4) \, , \\
p_1(\overline{E}) &= \frac{1}{\g^2}p_1(X_4) + \frac{1}{\pi^2}F\wedge F \, , &\qquad p_2(\overline{E}) &=  \frac{1}{\pi^2\g^2} p_1(X_4) \wedge F \wedge F \, .
\end{aligned}
\eeq
and substitution leads to 
\beq
\label{eq:SubLeadingS7}
\begin{split}
S_{7,{\rm top}} &= 2\pi \int_{X_4}p_1(X_4) \left( \frac{ 2 N^3_{M5} - N_{M_5} }{6} \frac{\g^2}{8\pi^2} \int_{Y^4} F \wedge F - \frac{N_{M_5}}{48} \int_{Y_4}p_1(Y_4) \right) \\
&= - cs(\internal) \left( \frac{ 2 N^3_{M5} - N_{M_5} }{3} \frac{\g^2}{8\pi^2} \int_{Y^4} F \wedge F - \frac{N_{M_5}}{24} \int_{Y_4}p_1(Y_4) \right) \, .
\end{split}
\eeq
These topological terms would appear in the semi-classical approximation of the gravity partition function. However, they are not well-defined even if $Y_4$ is a closed spin four-manifold. In this case, we know that $\g^2 \int F^2\in 8\pi^2\Z$, and $\int p_1(TY_4) = 3\sigma(Y_4) \in 48 \Z$ (by Rokhlin's theorem), so we can write
\beq
\begin{split}
S_{7,{\rm top}} &\in cs(\internal) \frac{ 2 N^3_{M5} - 7N_{M_5} }{3} \Z \, .
\end{split}
\eeq
For compact $\internal$, the Chern--Simons invariant is well-defined modulo $2\pi$, but the fraction is only an integer if $N_{M5}$ is a multiple of 3. This is consistent with the fact that the topological term in eleven dimensions is also not well-defined on its own \cite{Witten:1996md}. \\
The expression above also gives the subleading correction to \eqref{eq:UpliftM5} for the $\vartheta$ angle
\beq
\label{eq:ThetaCorrected}
\vartheta = cs(\internal) \frac{ 2 N^3_{M5} - N_{M_5} }{3} + o(N) \, .
\eeq
From the discussion in Section \ref{subsec:BdryTheta}, the presence of the finite counterterm \eqref{eq:FiniteCounterterm} guarantees that $\vartheta$ should be periodic with period $2\pi$ even on a space with a boundary. However, the expression above is not periodic unless $N_{M5}$ is a multiple of 3 (since $2n^3 + n\in 3\Z$ for any integer $n$). Thus, we expect other subleading terms to contribute. At the same order as \eqref{eq:S1loop} there are corrections to the Einstein equations of motion in the form of $R^4$ terms \cite{Tseytlin:2000sf}, but these reduce in seven dimensions to $R^3$ terms and thus should not contribute to the $\vartheta$ angle in the further compactification following the Pernici--Sezgin ansatz. 

The expression \eqref{eq:ThetaCorrected} has been obtained by reducing eleven-dimensional supergravity. However, it is also possible to obtain the higher-derivative correction to four-dimensional supergravity, as done in \cite{Bobev:2021oku}, and compare the results. Their expression for the Euclidean action including four-derivative corrections is
\beq
\begin{split}
S_{\rm HD} &= S + (c_1 + c_2) S_{W^2} + c_2 S_{\rm GB} + 16\pi^2 \ii c_3 \int p_1(Y_4) + 16 \ii (c_3 + c_4) \g^2 \int F \wedge F \, ,
\end{split}
\eeq
where $S$ is the two-derivative action \eqref{eq:4dSUGRAAction}, $S_{W^2}$ is a supersymmetrised version of the Weyl squared action, and $S_{\rm GB}$ is the Gauss--Bonnet action. This should be compared with the Wick-rotated version of the subleading action \eqref{eq:SubLeadingS7}, which leads us to conclude that, for this compactification
\beq
16\pi^2 c_3 = - cs(\internal) \frac{N}{24} \, , \qquad c_4 = 0 \, .
\eeq
These results have been obtained by looking at the subleading corrections directly in the eleven-dimensional effective Lagrangian. However, in this approach it is difficult to justify whether one has accounted for all the relevant terms, as there are ambiguities due to potential field redefinitions. So, it is important to check them using independent methods, such as amplitudes \cite{Chester:2018aca}.

\section{Field theory}
\label{sec:FieldTheory}

The AdS/CFT dictionary relates the on-shell gravity action in $d+1$ dimensions with the partition function of a field theory formulated on a $d$-dimensional manifold with the same geometric structure as the conformal boundary of the gravity solution. Which field theory should be considered depends on the knowledge of the full string/M-theory solution and hence on the choice of internal manifold.

\medskip

As showed in Section \ref{subsec:M2}, we can embed solutions of four-dimensional minimal supergravity in eleven-dimensional supergravity by uplifting on seven-dimensional Sasaki--Einstein manifolds. In this case, there is no $\vartheta$ term in gravity and the on-shell action is real. The dual field theories are known for numerous choices of Sasaki--Einstein manifolds and are generically non-Abelian Chern--Simons-matter theories, for which the localization procedure is often known and the large $N$ limit of the partition function successfully compared with the gravity observable. This is the case that was discussed in \cite{BenettiGenolini:2019jdz}.

On the other hand, in Section \ref{subsec:M5} it was shown that the consistent truncation of eleven-dimensional supergravity on $\internal\times S^4$ leads to an additional $\vartheta$ term in the four-dimensional supergravity action \eqref{eq:4dActionLorentzian}, which makes the on-shell action generically complex in Euclidean signature. The dual field theory on $M_3$ is a $\mc{N}=2$ SCFT that is obtained by wrapping $N$ $M5$-branes on $M_3\times \internal$ and then looking at the IR compared to the energies associated with the compact $\internal$ \cite{Gauntlett:2000ng}. It is the IR limit of a twisted compactification of the $A_{N-1}$ six-dimensional $(2,0)$ theory: the $SO(3)\subset SO(5)$ twisting in the gravity ansatz \eqref{eq:M5Ansatz} corresponds in the dual field theory to the topological twist by the $SO(3)_R\subset SO(5)_R$, which indeed leaves a commutant $SO(2)_R\cong U(1)_R$ symmetry dual to the gauge field in four-dimensional supergravity. This $R$-symmetry is the one used to couple to the supergravity background necessary to formulate the theory on a curved $M_3$. The investigation of the resulting field theory $T_N[\internal]$ goes under the name of \textit{$3d$-$3d$ correspondence}, and one of its crucial conjectures is that $T_N[\internal]$ should only depend on the topology of $\internal$ \cite{Dimofte:2011ju, Cecotti:2011iy, Dimofte:2011py}.

\medskip

The construction of $T_N[\internal]$ has not yet been fully untangled, one of the issues being whether the vacua of $T_N[\internal]$ account for all the $PSL(N;\C)$ flat connections on $\internal$ or only a subset thereof \cite{Chung:2014qpa}.\footnote{It is surely true that the structure of the theory is very different depending on whether $\internal$ is hyperbolic or not. In the latter case, if $\internal$ is a Seifert manifold the resulting field theory enjoys an additional $U(1)_f$ symmetry associated with the circle action of the Seifert fibration (see e.g. \cite{Gukov:2015sna, Pei:2015jsa, Gukov:2017kmk, Alday:2017yxk, Eckhard:2019jgg} for the construction).} Here we are interested in the case where $\internal$ is hyperbolic. From the supergravity viewpoint, this is necessary in the ansatz \eqref{eq:M5Ansatz}, as otherwise the resulting consistent truncation does not preserve supersymmetry or have an AdS vacuum \cite{Donos:2010ax}.

\medskip

Generically, the large $N$ limit of the free energy of the field theory is computed using a saddle point approximation, and the dominant saddles contributing to the evaluation could lead to a complex value for the free energy. This is known to be the case for partition functions corresponding to twisted indices: complex saddle points contribute in conjugate pairs resulting in an oscillating twisted index (there is a growing literature on the subject, see e.g. \cite{Benini:2018ywd, Choi:2019zpz, Cabo-Bizet:2019eaf, Choi:2019dfu, ArabiArdehali:2019orz, Murthy:2020rbd, Agarwal:2020zwm, Copetti:2020dil, Cabo-Bizet:2020ewf}). Here we have showed that, at least for theories arising from $M5$-branes, it is also true for the partition function on a generic fibred background. In fact, another crucial result of the $3d$-$3d$ correspondence is that the supersymmetric partition function of $T_N[\internal]$ on various spaces is captured by the partition function of $SL(N,\C)$ complex Chern--Simons theory on $\internal$ at an appropriate level (see \cite{Yagi:2013fda, Lee:2013ida, Cordova:2013cea, Dimofte:2014zga} for a derivation). Therefore, combining the expectations from AdS/CFT and $3d$-$3d$ correspondence provides us with a conjecture for the large $N$ limit for the partition function of complex Chern--Simons theory as well.

\medskip

To view concretely this limit, we first recall that Mostow's rigidity theorem guarantees that the hyperbolic metric on $\internal$ is unique, so ${\Vol}(\internal)$ is a topological invariant (in fact a homotopy invariant). The Chern--Simons invariant of the spin connection of the hyperbolic metric \eqref{eq:CSHyperbolic} is another topological invariant, which naturally combines with the hyperbolic volume in an invariant which is referred to as \textit{complex hyperbolic volume} \cite{Neumann:1985, Yoshida:1985}. More precisely, for a three-manifold both the dreibein and the spin connection can be seen as $\mf{so}(3)$ connections and we can introduce the $\mf{sl}(2,\C)$ geometric connection $\cA^{\rm geom} = \omega + \ii \e$ and its complex conjugate $\cA^{\overline{\rm geom}}$. This connection defines the complexified hyperbolic volume via the following relation
\beq
\label{eq:ComplexHyperbolicVolume_text}
\begin{split}
cs[\cA^{\rm geom}] &\equiv \frac{1}{8\pi}\int_{\internal} \tr_{\mb{2}} \left( \cA^{\rm geom} \wedge \rd \cA^{\rm geom} + \frac{2}{3} \cA^{\rm geom} \wedge \cA^{\rm geom} \wedge \cA^{\rm geom} \right) \\
&= \frac{\ii}{4\pi} \left({\Vol}(\internal) - \pi \ii \, {cs}(\internal)\right) \\
&\equiv \frac{\ii}{4\pi} \, {\Vol}_\C(\internal) \, , 	\\[5pt]
cs[\cA^{\overline{\rm geom}}] &= - \frac{\ii}{4\pi} \, {\Vol}_\C(\internal)^* \, .
\end{split}
\eeq
First, observe that these are precisely the combination appearing in \eqref{eq:IOS_IR} after substitution of \eqref{eq:UpliftM5}, so that we write
\beq
\label{eq:IIROnShell}
\begin{split}
I^{\rm IR} &= \frac{4\pi}{3\Delta\psi^2} N^3 \Bigg[ \sum_{\mathrm{nuts}_-} {\Vol}_\C(\internal)^* \frac{(b_1 + b_2)^2}{4b_1b_2} - \sum_{\mathrm{nuts}_+} {\Vol}_\C(\internal) \frac{(b_1 - b_2)^2}{4b_1b_2}\\
& \qquad \qquad \quad + \sum_{\rm bolts \ \Sigma_-} {\Vol}_\C(\internal)^* \int_{\Sigma_-} \left( \frac{1}{2}c_1(T\Sigma_-) + \frac{1}{4} c_1(N\Sigma_-) \right) \\
& \qquad \qquad \quad + \sum_{\rm bolts \ \Sigma_+} {\Vol}_\C(\internal) \int_{\Sigma_+} \left( \frac{1}{2}c_1(T\Sigma_+) - \frac{1}{4} c_1(N\Sigma_+) \right) \Bigg] \, .
\end{split}
\eeq
We see that the complexified hyperbolic volume appears naturally in the expression for the IR contribution to the on-shell action. Provided a smooth gravity solution exists, this expression gives the contributions of the saddle points to the large $N$ limit of the partition function of the field theory on $\bdry \cong \partial \bulk$ bounding $\bulk$. Of course, only some of them will be dominant. However, as already remarked, it is known that the field theory limit ``sees'' the contributions from various supergravity fillings, even in the case of Chern--Simons-matter field theories \cite{Toldo:2017qsh}.

\medskip

In order to be more concrete, we can consider various examples dual to those reviewed in Section \ref{sec:Examples}. To construct them, we shall need a few relations. First, given $h_1, h_2\in \mf{pgl}(2)$ and denoting by $\rho_N$ the $N$-dimensional irreducible representation of $\mf{pgl}(2)$, we have
\beq
\tr \left[ \rho_N(h_1) \rho_N (h_2) \right] = \frac{N^3-N}{6} \, \tr \left[ h_1 h_2 \right] \, .
\eeq
Secondly, we shall need the large $N$ expansion of the Reidemeister--Ray--Singer torsion \cite{Gang:2019uay}\footnote{This is the analytic torsion of an associated vector bundle in a representation and twisted by a flat connection. For the case relevant to us, we restrict to the adjoint $PSL(N,\C)$ bundle and the flat connection will be either $\cA^{\rm geom}$ or $\cA^{\overline{\rm geom}}$.}
\beq
\begin{split}
\mb{Tor}_{\rm adj}[\internal, \cA^{\rm geom}] &\to \exp \left( \frac{N^3}{3\pi}\Vol(\internal) + \ii \theta_{\internal, N} + \mc{O}(N) \right)  \, , \\
\mb{Tor}_{\rm adj}[\internal, \cA^{\overline{\rm geom}}] &\to \exp \left( \frac{N^3}{3\pi}\Vol(\internal) - \ii \theta_{\internal, N} + \mc{O}(N) \right) \, . 
\end{split}
\eeq
where $\theta_{\internal,N}$ is a real number. The large $N$ limit of the absolute value of the torsion is rigorously proved, whereas an expression for the phase factor has been conjectured in \cite{Choi:2020baw}, where numerical evidence has been gathered. Here we shall find additional evidence in support of the conjecture
\beq
\theta_{N,\internal} = - \frac{N^3}{3}cs(\internal) + o(N^3) \qquad \mod 2\pi \, ,
\eeq
to obtain
\beq
\begin{split}
\mb{Tor}_{\rm adj}[\internal, \cA^{\rm geom}] &\to \exp \left( \frac{N^3}{3\pi}\Vol_\C(\internal) + \mc{O}(N) \right)  \, , \\
\mb{Tor}_{\rm adj}[\internal, \cA^{\overline{\rm geom}}] &\to \exp \left( \frac{N^3}{3\pi}\Vol_\C(\internal)^* + \mc{O}(N) \right) \, . 
\end{split}
\eeq

\medskip

The solutions in Section \ref{subsec:SUSYBH} have trivial fibration and there is a non-trivial flux of the gauge field at the boundary, so they are dual to field theories on supersymmetric backgrounds $\bdry \cong S^1\times \Sigma_g$ where the $U(1)_R$ gauge field is used to implement a topological twist and the resulting partition function is a supersymmetric index topologically twisted by the $R$-symmetry \cite{Benini:2016hjo}. The requirement that $g>1$ for the supersymmetry of the gravity solutions corresponds to the fact that the large $N$ limit of the refined topologically twisted index on $S^1\times S^2$ vanishes (as showed using the $3d$-$3d$ correspondence in \cite{Benini:2019dyp}). For $g>1$, the twisted index of $T_N[\internal]$ on $S^1\times \Sigma_g$ is related to the Reidemeister--Ray--Singer torsion of the irreducible flat $SL(N,\C)$ connections on $\internal$. The large $N$ behaviour of the resulting invariant is given by \cite{Gang:2019uay}
\beq
\begin{split}
Z_{g,p=0} \left( T_N[\internal] \right) &\to N^{g-1} \mb{Tor}_{\rm adj}[\internal, \cA^{\rm geom}]^{g-1} + N^{g-1} \mb{Tor}_{\rm adj}[\internal, \cA^{\overline{\rm geom}}] \\
&\to \exp \left[ - \frac{N^3}{3\pi} \Vol_\C(\internal) (1-g) \right] + \exp \left[ - \frac{N^3}{3\pi} \Vol_\C(\internal)^* (1-g) \right] \, ,
\end{split}
\eeq
to leading order in $N$. The fact that modulus and phase combine into a complex topological invariant of $\internal$ was expected from the $3d$-$3d$ correspondence, since the partition function of $T_N[\internal]$ should only depend on the topology of $\internal$. The two contributions arise from saddle points corresponding to the geometric connection and its complex conjugate. As pointed out in \cite{Choi:2020baw}, the two terms in the sum can be explained by identifying each with one of the two limiting solutions in \eqref{eq:IOSBH}, or equivalently the two solutions \eqref{eq:IBoltIR}.
Having done this, we obtain
\beq
Z_{g,p=0} \left( T_N[\internal] \right) = \e^{- I[{\rm Bolt}_-]} + \e^{- I[{\rm Bolt}_+]} \, ,
\eeq
consistently with the expectations of AdS/CFT (we have set $\Delta\psi = 2\pi$ since both solutions are spin).

\medskip

This can be easily generalised to the case with non-trivial fibration. The partition function of $T_N[\internal]$ on the space $\mc{M}_{g,p\in 2\Z_{>0}}$ preserving two real supercharges can again be related to the Reidemeister--Ray--Singer torsion and the large $N$ of the resulting invariant computed \cite{Gang:2019uay}. In this case, the authors found a single saddle point that contribute to the evaluation, corresponding to the flat connection $\cA^{\overline{\rm geom}}$:
\beq
\begin{split}
Z_{g,p\in 2\Z_{>0}} \left( T_N[\internal] \right) &\to N^{g-1} \exp \left( 2p \ii  \, cs \left[ \rho_N \left( \cA^{\overline{\rm geom}} \right) \right] \right)  \mb{Tor}_{\rm adj}[\internal, \cA^{\overline{\rm geom}}]^{g-1} \\
&\to \exp \bigg[ - \frac{N^3}{3\pi} \Vol_\C(\internal)^* \left( 1-g-\frac{p}{4} \right) \bigg] \, .
\end{split}
\eeq
The dual gravity solution is the $\frac{1}{4}$-BPS solution with topology $\mc{O}(-p)\to \Sigma_g$ with action \eqref{eq:IBoltIR}, which is spin since we have chosen even $p$. So we write
\beq
Z_{g,p\in 2\Z_{>0}} \left( T_N[\internal] \right) \to \e^{- I^{\rm IR} \left[ {\rm Bolt}_+ \right]} \, .
\eeq
The existence of a unique saddle point is due to the fact that the family Bolt$_-$ has larger on-shell action for $p>0$, whereas when $p=0$ the two solutions have the same on-shell action and thus have the same claim to represent a saddle point of the gravitational path integral. Furthermore, notice that the matching with the field theory supersymmetric partition function doesn't require $I^{\rm UV}$ \eqref{eq:IBoltUV}, thus providing further evidence in favour of the necessity of the finite counterterm \eqref{eq:FiniteCounterterm}.

\medskip

Finally, another supersymmetric background important for the $3d$-$3d$ correspondence is the $U(1)\times U(1)$ squashed three-sphere, which preserves two supercharges \eqref{eq:SquashedU1U1}. The partition function of $T_N[\internal]$ on this background should be equal to the partition function of $SL(N,\C)$ on $\internal$ with parameters $\hbar$ and $\tilde{\hbar}$ related to the squashing
\beq
Z_b\left( T_N[\internal] \right) = Z_{\internal} \left( CS; \hbar = 2\pi \ii b^2, \tilde{\hbar} = 2\pi \ii b^{-2} \right) \, .
\eeq
The observable on the right-hand side is hard to compute. However, it was shown in \cite{Gang:2014ema} that the non-perturbative effects of order $\e^{-4\pi^2/\hbar}$ are subleading in $N$, at least for the absolute value. If we can ignore the non-perturbative corrections, then $Z_{\Sigma_3}(CS)$ has the same asymptotic expansion in $\hbar$ as the perturbative expansion of the $SL(N,\C)$ Chern--Simons partition function around the saddle point $\cA^{\overline{\rm geom}}$:
\beq
\begin{split}
Z_{\internal} \left( CS; \hbar, \tilde{\hbar} \right) &\sim Z^{\rm pert}_{\internal} \left( CS; \hbar, \tilde{\hbar}; \cA^{\overline{\rm geom}} \right)  \\
&= \exp \left( \frac{1}{\hbar} S_0^{\cA^{\overline{\rm geom}}} - \frac{3}{2} \log \hbar + S_1^{\cA^{\overline{\rm geom}}}  + \hbar S_2 ^{\cA^{\overline{\rm geom}}}  + \dots + \hbar^{n-1} S_n^{\cA^{\overline{\rm geom}}}  + \dots \right)
\end{split}
\eeq
with
\beq
\begin{split}
S_0^{\cA^{\overline{\rm geom}}} &= 4\pi \, cs \left[ \rho_N \left( \cA^{\overline{\rm geom}} \right) \right] \sim - \ii \frac{N^3}{6}\Vol_\C(\internal)^* + o(N^3) \, , \\
S_1^{\cA^{\overline{\rm geom}}} &= - \frac{1}{2} \log \mb{Tor}_{\rm adj}[\internal, \cA^{\overline{\rm geom}}] \sim - \frac{N^3}{6\pi} \Vol_\C(\internal)^* + o(N^3) \, .
\end{split}
\eeq
The expression for $S_n$ with $n\geq 2$ are not proved rigorously. However, the authors of \cite{Gang:2014ema} found numerical evidence in support of the following conjecture
\beq
\begin{split}
\lim_{N\to \infty} \frac{1}{N^3}\Im \left[ S_2^{\cA^{\overline{\rm geom}}} \right] &= \frac{1}{24\pi^2} \Vol(\internal) \, , \\
\lim_{N\to \infty} \frac{1}{N^3}\Im \left[ S_{2j}^{\cA^{\overline{\rm geom}}} \right] &= \lim_{N\to \infty} \frac{1}{N^3}\Re \left[ S_{2j-1}^{\cA^{\overline{\rm geom}}} \right] = 0 \, .
\end{split}
\eeq
\beq
\begin{split}
\lim_{N\to \infty} \frac{1}{N^3} S_2^{\cA^{\overline{\rm geom}}} &= \frac{\ii}{24\pi^2} \Vol_\C(\internal)^* \, , \\
\lim_{N\to \infty} \frac{1}{N^3} S_{n}^{\cA^{\overline{\rm geom}}} &= 0 \mod 2\pi \ii \qquad n \geq 3 \, .
\end{split}
\eeq
According to this conjecture, we would find
\beq
\begin{split}
Z_b \left( T_N[\internal] \right) &\to \exp \bigg[ -  \frac{N^3}{3\pi}\frac{1}{4} \left( b + \frac{1}{b} \right)^2 \Vol_\C(\internal)^* \bigg] \, . 
\end{split}
\eeq
The dual solution is AdS$_4^{(b)}$, meaning AdS$_4$ with an instanton considered in Section \ref{subsec:AdS4}, with IR action \eqref{eq:IAdSIR}, and indeed
\beq
Z_b \left( T_N[\internal] \right) \to \e^{- I^{\rm IR}[ {\rm AdS}_4^{(b)}] } \, ,
\eeq
Again the presence of a single saddle point corresponds to the gravity solution with larger on-shell action, and the absence of $I^{\rm UV}$ \eqref{eq:IAdSUV} supports the finite counterterm \eqref{eq:FiniteCounterterm}.

\medskip

As far as we know, this list exhausts the supersymmetric backgrounds for which the large $N$ limit of the $3d$-$3d$ correspondence has been studied. However, the examples considered in Section \ref{sec:Examples} allow us to immediately conjecture the values of additional observables in the large $N$ limit. For instance, \eqref{eq:IBoltIR} shows that the large $N$ limit of the partition function of $T_N[\internal]$ on squashed $\cM_{g,p}$ should not depend on the squashing parameter, consistently with the standard analysis of \cite{Closset:2013vra}. Similarly, the analysis of Section \ref{subsec:OtherTopologies} gives an expectation for the value of the large $N$ limit of the partition function of $T_N[\internal]$ on squashed Lens spaces, as studied in \cite{Dimofte:2014zga}. The interpretation of these partition functions is not obvious from the $3d$-$3d$ correspondence.

We should also point out that the supergravity ansatz made in order to construct the four-dimensional solutions holographically dual to $T_N[\internal]$ requires $\internal$ to be compact. However, it is also possible to construct a field theory $T_N[\internal]$ for \textit{cusped} $\internal$, meaning that it is non-compact but with finite volume. In this case, it is more difficult to define a Chern--Simons invariant because it is more difficult to find a section of the frame bundle. It is still possible to define an analogous invariant \cite{Meyerhoff}, in which case the complex hyperbolic volume is defined modulo $\ii\pi^2$ and still related to the integral of a $\mf{sl}(2,\C)$-valued geometric connection \cite{Yoshida:1985}. This invariant, though, also includes contributions from the cusps, which is not clear how to introduce in the supergravity setup.

\section*{Acknowledgments}

\noindent
I am grateful to James Sparks for very helpful discussions and comments on the draft.
I have also benefited from conversations with Francesco Benini, Davide Cassani, Jerome Gauntlett, Chiung Hwang, Paul Richmond, Luigi Tizzano, David Tong, and Carl Turner. I would also like to thank the anonymous referee for their helpful suggestions to improve the manuscript. My work has been supported by the Simons Foundation, by the STFC consolidated grant ST/T000694/1, and by the ERC Consolidator Grant N. 681908 ``Quantum black holes: A macroscopic window into the microstructure of gravity.'' I also gratefully acknowledge support and hospitality from the Galileo Galilei Institute.

\appendix

\section{Chern--Simons conventions}
\label{sec:CSInvariant}

Let $A=A^\alpha T^\alpha$ be a $\g$ connection, and $T^\alpha$ the basis of $\g$, with the following normalizations
\beq
\begin{split}
\tr_{\mb{r}}(T^\alpha T^\beta) &= C(\mb{r}) \, \delta^{\alpha\beta} \, , \qquad [T^\alpha, T^\beta] = f^{\alpha\beta\gamma} T^\gamma \, , \\
F &= \rd A + A \wedge A = \left( \rd A^\alpha + \frac{1}{2}f^{\alpha\beta\gamma}A^\beta \wedge A^\gamma \right) T^\alpha \, ,
\end{split}
\eeq
where $C(\mb{r})$ is (minus) the index of the representation $\mb{r}$. The Chern--Simons 3-form is defined by
\beq
\label{eq:CSGeneral}
\begin{split}
Q_3 [A;\g,\mb{r}] &\equiv {\tr_{\mb{r}}} \left( A \wedge \rd A + \frac{2}{3} A \wedge A \wedge A \right) \\
&= C(\mb{r}) \left( A^\alpha \wedge \rd A^\alpha + \frac{1}{3}f^{\alpha\beta\gamma} \, A^\alpha \wedge A^\beta \wedge A^\gamma \right) \, .
\end{split}
\eeq
It has the property that $\rd Q_3 [A;\g,\mb{r}] = {\tr_\mb{r}} ( F \wedge F)$.\\
Now extend the definition to a complexified gauge group $G_\C$ with algebra $\g_\C$ for which $G$ is the compact real form. Write $\cA \equiv A + \ii B$ for the $\g_\C$-valued connection with $A, B$ $\g$-valued connections. The curvature is
\beq
\label{eq:ComplexLieAlgebraCurvature}
\cF = F_A + \ii \, D_AB - B \wedge B \, ,
\eeq 
where
\beq
(D_AB)^\alpha \equiv \rd B^\alpha + f^{\alpha\beta\gamma} A^\beta \wedge B^\gamma \, .
\eeq
The Chern--Simons form for the complex connection is
\beq
\label{eq:Q3ComplexGroup}
\begin{split}
Q_3[\cA; \g_\C,\mb{r}] &= \tr_{\mb{r}}\left( \cA \wedge \rd \cA + \frac{2}{3} \cA \wedge \cA \wedge \cA \right) \\
&= Q_3[A; \g,\mb{r}] - \tr \left( B \wedge D_A B \right) + 2\ii \, \tr_{\mb{r}} \left( B \wedge F_A - \frac{1}{3}B \wedge B \wedge B \right) \\
& \ \ \ - \ii \, \rd ( \tr_{\mb{r}} A \wedge B )
\end{split}
\eeq
The action of $G_\C$ Chern--Simons theory is
\beq
\label{eq:ICSGC}
\begin{split}
I &= \frac{k+\ii s}{8\pi}\int Q_3[\cA;\g, \mb{r}] + \frac{k-\ii s}{8\pi}\int Q_3[\overline{\cA};\g, \mb{r}] \\
&= \frac{k}{4\pi}\int \left( Q_3[A; \g,\mb{r}] - \tr \left( B \wedge D_A B \right) \right) - \frac{s}{2\pi}\int \tr_{\mb{r}} \left( B \wedge F_A - \frac{1}{3}B \wedge B \wedge B \right) \, .
\end{split}
\eeq
Invariance under large gauge transformations requires $k\in\Z$, and $u$ is either real or purely imaginary in order to have a unitary theory \cite{Witten:1989ip}.

\medskip

We are interested in $\mf{so}(N)$ bundles. A connection of $\mathfrak{so}(N)$ in the defining representation is concretely represented by an anti-symmetric matrix $A = A^{ab}$ with $A^{ab}=-A^{ba}$. Then we can define the Chern--Simons form (for $\mathfrak{so}(N)$, $C(\mb{fund}) = 1$)
\beq
Q_3[A; \mathfrak{so}(N), \mb{fund}] = \tr \left( A \wedge \rd A + \frac{2}{3} A \wedge A \wedge A \right) \, .
\eeq
A standard argument based on extension in the bulk then shows that the integral of $Q_3[A; \mathfrak{so}(N), \mb{fund}]$ on $\internal$ is well-defined modulo $16\pi^2$ if we choose a spin structure on $\internal$, otherwise modulo $8\pi^2$. \\
Over a three-manifold there is a special $\mf{so}(3)$ bundle, the frame bundle. The connection is the spin connection $\omega^{ab}$, and we can define the Chern--Simons form
\beq
\label{eq:CSOmega3}
\begin{split}
Q_3[\omega;\mf{so}(3),\mb{3}] &\equiv \tr_{\mb{3}} \left( \omega \wedge \rd\omega + \frac{2}{3}\omega\wedge \omega \wedge \omega \right) \, .
\end{split}
\eeq
For a closed oriented 3-manifold with a choice of spin structure, this defines a topological invariant defined modulo $2\pi$
\beq
\label{eq:CS3Normalised_Spun}
cs(\internal) \equiv \frac{1}{8\pi}\int_{\internal}Q_3[\omega;\mf{so}(3),\mb{3}] \in \R/2\pi\Z \, ,
\eeq
which is well-defined modulo $2\pi$. However, it is also possible to define a topological invariant without choosing the spin structure, using the Chern--Simons ``at level one'' \eqref{eq:ChernSimonsLevelOne}
\beq
\frac{1}{4\pi}\int_{\internal}Q_3[\omega;\mf{so}(3),\mb{3}] \, .
\eeq

\medskip

In fact, for $\mf{so}(3)$ more is true, since both the dreibein and the spin connection can be seen as $\mathfrak{so}(3)$ connections. Thus, we can dualize the index and write
\beq
\widetilde{\omega}^a \equiv \frac{1}{2}\epsilon^{abc}\omega^{bc} \, , \qquad \omega^{ab} = \epsilon^{abc}\widetilde{\omega}^{c} \, .
\eeq
dualizing the index. This allows us to construct a natural complex connection for $\g_\C = \mf{sl}(2,\C)$:
\beq
\cA_{\rm geom} \equiv \widetilde{\omega}^a + \ii \, \e^a \, .
\eeq
This is the \textit{geometric connection}, whose curvature can be found applying \eqref{eq:ComplexLieAlgebraCurvature} 
\beq
\label{eq:Fgeom}
\begin{split}
{\cF}_{\rm geom}^a &= \widetilde{\rho}^a + \frac{1}{2}\epsilon^{abc} e^b \wedge e^c \, ,
\end{split}
\eeq
where $\rho^{ab}$ is the curvature two-form.
Substitution in \eqref{eq:Q3ComplexGroup} leads to
\beq
\begin{split}
\int Q_3[\cA_{\rm geom} ; \mf{sl}(2,\C),\mb{2}] &= \int Q_3[\omega; \mf{su}(2),\mb{2}] + 2\ii \int \tr_{\mb{2}} \left( \e \wedge \widetilde{\rho} - \frac{1}{3} \e \wedge \e \wedge \e \right) \\
&= \int Q_3[\omega; \mf{su}(2),\mb{2}] - \frac{\ii}{2} \int \left( R  + 2 \right)  \vol_g
\end{split}
\eeq
Changing the representation of $\mf{su}(2)$, we conclude that for a compact 3-manifold with a chosen spin structure
\beq
\begin{split}
\label{eq:ExpansionCS_1}
\int_{\internal} Q_3[\cA_{\rm geom} ; \mf{sl}(2,\C),\mb{2}] &= 2\pi \, cs(\internal) - \frac{\ii}{2} \int_{\internal} \left( R  + 2 \right)  \vol_g
\end{split}
\eeq

\medskip

Now choose a hyperbolic 3-manifold. By hyperbolic 3-manifold we mean a complete Riemannian three-dimensional manifold with constant sectional curvature $-1$. All these manifolds can be obtained as quotients of $\mathbb{H}^3/\Gamma$, where $\Gamma$ is a \textit{Kleinian group}, a discrete subgroup of $PSL(2,\C)$. We also restrict to the case of manifolds with finite volume. Mostow's rigidity theorem guarantees that for this class of manifolds the hyperbolic structure is uniquely determined by the homotopy type, so the volume is a topological invariant. Note that these manifolds (with finite volume) can be compact or non-compact. A non-compact hyperbolic manifold without boundary and finite volume is also called \textit{cusped}.

For a hyperbolic manifold, \eqref{eq:Fgeom} shows that the geometric connection is \textit{flat}, and we can simplify \eqref{eq:ExpansionCS_1} to
\beq
\int_{\internal} \left( - \frac{\ii}{2} Q_3[\cA_{\rm geom} ; \mf{sl}(2,\C),\mb{2}] \right) = {\rm Vol}(\internal) - \pi \ii \, cs(\internal) \, .
\eeq
The quantity
\beq
\label{eq:ComplexHyperbolicVolume}
\Vol_\C(\internal) \equiv \Vol(\internal) - \pi \ii \, cs(\internal) = \int_{\internal} \left( - \frac{\ii}{2}Q_3[\cA_{\rm geom} ; \mf{sl}(2,\C),\mb{2}]  \right)
\eeq
is the \textit{complex hyperbolic volume} of the compact hyperbolic three-manifold \cite{Neumann:1985, Yoshida:1985}. It is a topological invariant defined modulo $\ii 2\pi^2$ (having chosen the spin structure). Therefore, it is natural to write it as an exponential
\beq
\exp \left( \frac{1}{\pi} \Vol_\C(\internal) \right) = \exp \left( \frac{1}{\pi} \Vol_\C(\internal) \right) \e^{ - \ii \, cs(\internal) } \, .
\eeq

\medskip

We are also interested in non-compact hyperbolic three-manifold (that is, with cusps). For instance, most knot complements are non-compact. In this case, it not necessarily true that there is a global section of the frame bundle that can be used to compute the Chern--Simons invariant using the formula \eqref{eq:CSOmega3} with $Q_3$. However, Meyerhoff \cite{Meyerhoff} has showed that there is a way of defining a Chern--Simons invariant by defining a special singular frame, and adding the contribution from the singularities at the cusps:
\beq
\begin{split}
cs^{\rm Me}(\internal) &= - \frac{1}{8\pi}\int_{s(\internal \setminus L)} Q_3[\omega; \mf{so}(3),\mb{3}] + \frac{1}{2} \sum_{K\in L} \tau(K) \mod \pi \, , 
\end{split}
\eeq
where $L$ is a link in $\internal$, with components $K$, $s:\internal \setminus L \to F(M_3)$ is a special singular frame, and $\tau(K)$ is the torsion of the singular curve $K$ in $L$. This can also be used to construct a complexification of the hyperbolic volume, by defining
\beq
\label{eq:ComplexHyperbolicVolumeMeyerhoff}
\Vol_\C^{\rm Me}(\internal) \equiv \Vol(\internal) - \pi \ii \, cs^{\rm Me}(\internal) \, ,
\eeq
which is defined modulo $\ii \pi^2$. Just like in the case of compact hyperbolic manifolds, this Chern--Simons invariant can also be related to the integral of the $\mf{sl}(2,\C)$-valued geometric connection \cite{Yoshida:1985}.

\section{Bott--Cattaneo formula}
\label{app:AngularForms}

For an $\mf{o}(N)$ vector bundle $E$ with curvature $F$ (represented by an antisymmetric matrix) we can define the Pontryagin classes
\beq
\label{eq:Pontryagin}
p_1 \equiv - \frac{1}{8\pi^2}\tr \, F^2 \, , \qquad p_2 \equiv \frac{1}{8(2\pi)^4} \left[ (\tr \, F^2 )^2 - 2 \, \tr \, F^4 \right] \, .
\eeq
This is in particular true of the tangent bundle.

\medskip

Let $E\to B$ be a rank $2m+1$ vector bundle over $B$. The fiber over $p\in B$ is a copy of $\R^{2m+1}$ parametrized by $y^a$, $a=1, \dots, 2m+1$, equipped with a fiber metric. The \textit{sphere bundle} $S(E)$ associated to $E$ is the bundle $S(E)\to B$ whose fiber over $p\in B$ is the unit sphere $S^{2m}\subset E_p \cong \R^{2m+1}$. This bundle has structure group $SO(2m+1)$ and the non-triviality is represented by the connection $A^{ab}$. This allows us to define the forms
\beq
D y^a \equiv \rd y^a + \g A^{ab}y^b \, , \qquad F^{ab} \equiv \rd A^{ab} + \g A^{ac}\wedge A^{cb} \, .
\eeq
The \textit{global angular form} for $S(E)$ is the unique closed and gauge-invariant improvement of the sphere volume form. For $m=1, 2$, it has the following form
\begin{align}
\label{eq:E2}
\E_2 &\equiv \frac{1}{8\pi} \epsilon_{a_1a_2a_3} \, y^{a_1} \left[ D y^{a_2}\wedge D y^{a_3} + \g \, F^{a_2a_3} \right] \, , \\
\label{eq:E4}
\begin{split}
\E_4 &\equiv \frac{1}{64\pi^2} \epsilon_{a_1a_2a_3a_4a_5} \, y^{a_1} \Big[ D y^{a_2}\wedge D y^{a_3} \wedge D y^{a_4} \wedge D y^{a_5} + 2\g \, F^{a_2a_3} \wedge Dy^{a_4} \wedge D y^{a_5} \\
& \qquad \qquad \qquad \qquad \qquad + \g^2 \, F^{a_2a_3}\wedge F^{a_4a_5} \Big] \, .
\end{split}
\end{align}
Here, we have chosen a normalization such that the global angular form integrates to $1$ along the fiber:
\beq
\int_{S^{2m}} \E_{2m} = 1 \, .
\eeq
The Bott--Cattaneo formula gives a relation between the integral of the global angular form along the fibers, and the characteristic classes of the $SO(2m+1)$ bundle over $B$:
\beq
\label{eq:BottCattaneo}
\int_{S^{2m}} \left( \E_{2m} \right)^{2s+2} = 0 \, , \qquad \int_{S^{2m}}\left( \E_{2m} \right)^{2s+1} = \frac{\g^{2ms}}{2^{2s}}\left[ p_m(E) \right]^s \, .
\eeq

\bibliographystyle{./ThetaFiles/JHEP}
{\small
\bibliography{./ThetaFiles/Bib_Theta}}

\end{document}